\documentstyle[prl,aps,preprint,amsfonts]{revtex}
\newcommand{\tr}{{\rm tr}}
\newcommand{\Tr}{{\rm Tr}}
\newcommand{\Det}{{\rm Det}}
\newcommand{\Dh}{{\cal D}}
\newcommand{\eq}[1]{eq.~(\ref{eq:#1})}
\newcommand{\bfa}{\mbox{\boldmath $a$}}
\newcommand{\bfp}{\mbox{\boldmath $p$}}
\newcommand{\bfx}{\mbox{\boldmath $x$}}
\newcommand{\bfA}{\mbox{\boldmath $A$}}
\newcommand{\bfD}{\mbox{\boldmath $D$}}
\newcommand{\bfE}{\mbox{\boldmath $E$}}
\newcommand{\bftau}{\mbox{\boldmath $\tau$}}
\newcommand{\bfnabla}{\mbox{\boldmath $\nabla$}}
\newcommand{\Li}{{\rm Li}}
\newcommand{\bfAr}{\mbox{\boldmath ${\cal A}$}}

\begin{document}

\draft
\tighten

\title{Gauge invariant derivative expansion of the effective action at
finite temperature and density and the scalar field in 2+1
dimensions }

\author{C. Garc\'{\i}a-Recio and L.L. Salcedo}

\address{
{~} \\
Departamento de F\'{\i}sica Moderna \\
Universidad de Granada \\
E-18071 Granada, Spain
}

\date{\today}
\maketitle

\thispagestyle{empty}

\begin{abstract}
A method is presented for the computation of the one-loop effective
action at finite temperature and density. The method is based on an
expansion in the number of spatial covariant derivatives. It applies
to general background field configurations with arbitrary internal
symmetry group and space-time dependence. Full invariance under small
and large gauge transformations is preserved without assuming
stationary or Abelian fields nor fixing the gauge. The method is
applied to the computation of the effective action of spin zero
particles in 2+1 dimensions at finite temperature and density and in
presence of background gauge fields. The calculation is carried out
through second order in the number of spatial covariant
derivatives. Some limiting cases are worked out.
\end{abstract}


\section{Introduction}\label{sec:1}

In their pioneering work, Deser, Jackiw and Templeton
\cite{Deser:1982vy} noted that gauge theories in odd-dimensional
spaces naturally admit a local term of topological nature, known as
Chern-Simons term\cite{Chern:1974} (see \cite{Dunne:1998qy} for a
recent review). One of the interesting properties of the non-Abelian
Chern-Simons term is that under gauge transformations it changes
proportionally to the winding number of the transformation. Thus, when
the action contains a Chern-Simons term, the partition functional of
the system is only well defined if the coupling constant of the
Chern-Simons term is properly quantized.

It was later realized
\cite{Redlich:1984kn,Niemi:1983rq,Alvarez-Gaume:1984ig} that the
Chern-Simons term is induced by quantum fluctuations when gauge fields
are coupled to odd-dimensional fermions. Such term comes out with the
correctly quantized coupling constant and so full gauge invariance is
preserved (although possibly at the price of spoiling parity
invariance\cite{Redlich:1984kn}). Because the Chern-Simons term is a
polynomial in the gauge fields and their derivatives, it can be
obtained through a combination of perturbative and derivative
expansions.

In the so-called imaginary time formalism for field theory at finite
temperature\cite{Matsubara:1955ws,Landsman:1987uw}, the space-time has
a non-trivial topology, since the time is effectively compactified to
a circle. This allows the existence of topologically large gauge
transformations even in the Abelian case. When the problem of the
induced Chern-Simons term is studied using the method just mentioned
of retaining a low number of fields and of derivatives, a puzzling
situation appears, namely, the coefficient of the Chern-Simons term
turns out to be a smooth function of the temperature, and hence it
violates the quantization condition\cite{Ishikawa}. The situation has
recently been clarified by considering a simple $0+1$-dimensional
model\cite{Dunne:1997yb} which can be computed in closed form. There
it is seen that full gauge invariance holds for the exact result but
it is broken by perturbation theory. This is not difficult to
understand, since in simple cases gauge invariance under large gauge
transformations is equivalent to periodicity of the effective action
as a function of the gauge field, whereas perturbation theory
corresponds to a Taylor expansion of that function. Clearly, the
property of being periodic is not maintained in general by a truncated
Taylor expansion. In \cite{Deser:1997nv} it was noted that full gauge
invariance is always a property of the exact result, since it follows
straightforwardly from using a $\zeta$-function regularization. In
\cite{Fosco:1997ei} the exact effective action of fermions in $2+1$
dimensions was obtained for the case of Abelian and stationary
background gauge fields.

The problem of preserving full gauge invariance at finite temperature
is not tied to odd-dimensional theories
\cite{Deser:1998gp,Das:1999gc,Fosco:1999md,Dunne:2000ek,Salcedo:1998tg}
nor to fermions \cite{Barcelos-Neto:1999zv,Das:2000rc}. It appears
whenever perturbation theory is involved. This is unfortunate, since,
as noted in \cite{Dunne:1998qy}, ``at finite temperature, perturbation
theory is one of the few tools we have.'' In this work we show that it
is possible to carry out detailed calculations of the effective action
fully preserving gauge invariance, without restricting oneself to
particular configurations such as Abelian or stationary ones, and
without choosing a particular gauge. The study of simple cases
\cite{Dunne:1997yb,Deser:1997nv,Fosco:1997ei} shows that the problem
with gauge invariance comes through the scalar potential $A_0(x)$.
The finite temperature effective action is non-local in time but it is
local in the space variables. This suggests to consider an expansion
in the number of spatial covariant derivatives only. The time
component is treated non-perturbatively in order to avoid destroying
gauge invariance. (See \cite{Barcelos-Neto:1998xh} for another
discussion of derivative expansions at finite temperature.)

It should be emphasized that the expansion in the number of spatial
covariant derivatives is not tied to a particular method of
computation, since it can be obtained from the exact result (namely,
by considering an appropriate spatial dilatation of the background
fields) and is fully gauge invariant. What it is shown here is that it
is also amenable to explicit computation order by order, through a
combination of the method of symbols and $\zeta$-function, for
instance. This combination works very well and has been applied at
zero temperature for fermions with local\cite{Salcedo:1996qy} and
non-local actions\cite{RuizArriola:1999zi}. At finite temperature it
has been applied to fermions in odd-dimensions\cite{Salcedo:1999sv} as
well as in even-dimensions\cite{Salcedo:1998tg}, however, for
technical reasons, this has been done choosing a particular gauge. In
the present work we remove the necessity of any choice of gauge. It
turns out that previous formulas in
\cite{Salcedo:1999sv,Salcedo:1998tg} can be reinterpreted and
rewritten in a manifest gauge invariant form.

Although a naive perturbative expansion in $A_0$ breaks gauge
invariance, the very method of calculation suggests an expansion in
powers of the temporal covariant derivative in the adjoint
representation which preserves gauge invariance. This yields the
remarkable result that, even at finite temperature, the theory is
local if expressed in the appropriate variables.

The method is explicitly applied to the case of relativistic scalar
particles in $2+1$ dimensions. The computation is carried out through
second order in the number of spatial covariant derivatives. Contact
is made with the relativistic Bose gas.

\section{Gauge invariant derivative expansion at finite temperature}
\label{sec:2a}
\subsection{The mathematical problem}

The aim of this section is to present a scheme to address the
computation of the one-loop effective action at finite temperature
preserving gauge invariance at every step. (We note that gauge
invariance in this work refers always to vector gauge
transformations.) To fix ideas consider the case of scalar particles
in $d+1$ dimensions in presence of background gauge fields. This case
will be worked out later for $d=2$. The Euclidean action of the system
is
\begin{equation}
S= 
\int d^{d+1}x\left((D_\mu\phi)^\dagger(D_\mu\phi)+m^2\phi^\dagger\phi\right)\,
\label{eq:n1}
\end{equation}
where $D_\mu=\partial_\mu+A_\mu$ is the covariant derivative. The
finite temperature condition can be implemented by using the imaginary
time formalism, that is, by compactification of the Euclidean time to
a circle so that the fields $\phi$ and $A_\mu$ are periodic functions
of $x_0$ with period $\beta=1/T$ ($T$ being the temperature). After
functional integration over $\phi(x)$, the Euclidean effective action
is formally given by
\begin{equation}
W_s[m,A]= \Tr_b\log(-D_\mu^2+m^2)\,.
\label{eq:n2}
\end{equation}
The subindex $b$ recalls that the functional trace is to be taken in
the Hilbert space of bosonic wave functions, i.e. with periodic
boundary conditions.

Presently the mathematical problem to be addressed is the computation
of quantities of the form
\begin{equation}
\Gamma[M,A]= \Tr(f(M,D))
\end{equation}
where $D_\mu$ is the covariant derivative and $M(x)$ collectively
denote one or more matrix valued functions of $x_\mu$ representing
other external fields in addition to the gauge fields. The trace
refers to the Hilbert space $\cal{H}$ of wavefunctions with space-time
and internal degrees of freedom, the space-time manifold has topology
${\cal M}_{d+1}={\rm S}^1\times{\cal M}_d$ and the wavefunctions are
periodic for bosons and antiperiodic for fermions.

Although slightly pedantic, it will occasionally be convenient to
regard $M$ not as functions but rather as multiplicative operators
in ${\cal H}$, i.e., operators commuting with the operators $x_\mu$
and otherwise with arbitrary structure in internal space. Likewise
$D_\mu$ are differential operators of the form $\partial_\mu+A_\mu$
with $A_\mu$ multiplicative. The quantity $f(M,D)$ denotes an operator
constructed out of $M$ and $D_\mu$ in the algebraic sense, that is,
$f(M,D)$ is a linear combination (or series) of products of $M$ and
$D_\mu$ multiplied in any order with constant c-number
coefficients. In order for $M$ and $D_\mu$ to be well-defined
operators in ${\cal H}$, $M(x)$ and $A_\mu(x)$ are required to be
periodic functions of $x_0$. In addition we will assume that the
fields are sufficiently convergent at infinity and the function $f$ is
well-behaved. This means, in particular, that $f$ is one-valued and
sufficiently convergent at infinity as a function of $D_\mu$ to ensure
the existence of the trace (by avoiding ultraviolet divergences).

A gauge transformed configuration $(M^U,A^U)$ is one of the form
\begin{equation}
M^U(x)=U^{-1}(x)M(x)U(x)\,,\quad
A^U_\mu(x)= U^{-1}(x)\partial_\mu U(x) +U^{-1}(x)A_\mu(x)U(x)
\,,
\end{equation}
where the gauge transformation $U(x)$ is a periodic function of $x_0$
which takes values on matrices in internal space.  This corresponds to
a similarity transformation of $D_\mu$, namely, $D^U_\mu=
\partial_\mu+A^U_\mu = U^{-1}D_\mu U$ where $U$ is to be regarded as a
multiplicative operator in $\cal{H}$.  Because $f(M,D)$ is constructed
with $M$, $D$ and c-numbers, it follows that $f(M,D)$ also transforms
under a similarity transformation
\begin{equation}
f(M^U,D^U)= U^{-1}f(M,D)U
\end{equation}
and so 
\begin{equation}
\Gamma[M^U,A^U]= \Gamma[M,A]
\end{equation}
using the cyclic property of the trace, which holds due to our
regularity assumptions for $f(M,D)$.\footnote{In practice,
$\Gamma[M,A]$ is only computed for a subset of configurations
$(M,A)$ and only the subgroup of gauge transformations which leave
invariant such a subset are relevant. For fermions, the internal space
includes Dirac space as well flavor degrees of freedom, and the
$\gamma_\mu$ matrices are included in $M$ (they are not c-numbers). In
this case only gauge transformations in flavor space are relevant
since they are the ones that preserve the form of $\gamma_\mu$.}

\subsection{The method of symbols}

Assuming that the operator $\hat{f}=f(M,D)$ admits a complete set of
eigenfunctions, $\hat{f}|n\rangle =\lambda_n|n\rangle$, the functional
trace is simply $\Gamma[M,A]=\sum_n\lambda_n$. In this form gauge
invariance is obvious since $\hat{f}$ and $\hat{f}^U$ are related by a
similarity transformations and hence they have the same spectrum.

The gauge invariance of $\Gamma[M,A]$ is also manifest computing the
trace in the basis $|x\rangle$ of eigenfunctions of $x_\mu$,
normalized as $\langle x| x^\prime\rangle = \delta(x-x^\prime)$ (a
periodic delta function in the temporal direction)
\begin{equation}
\Gamma[M,A]= \int d^{d+1}x\,\tr\langle x|f(M,D)|x\rangle
\end{equation}
(where $\tr$ refers to internal space) because $U$ is a multiplicative
operator and so $\tr\langle x|f(M,D)|x\rangle$ is gauge invariant
without integration over $x$. However, computationally it is more
convenient to use a basis in momentum space $|p\rangle$
\begin{equation}
\langle x|p\rangle= e^{px}\,, \quad
\langle p|p^\prime\rangle=
\beta \delta_{p_0 p^\prime_0}(2\pi)^d\delta(\bfp-\bfp^\prime)\,,
\end{equation}
(to avoid unessential factors of $i$ we take the convention of using
purely imaginary momenta $p_\mu$ but $\int d^d\bfp$ below denotes the
usual integral in ${\mathbb{R}}^d$ and $\delta(\bfp-\bfp^\prime)$ denotes
the corresponding delta function). The frequency takes the Matsubara
values $p_0= 2\pi in/\beta$ for bosons and $p_0= 2\pi
i(n+\frac{1}{2})/\beta$ for fermions. Note that we have assumed that
the space manifold ${\cal M}_d$ has a topology ${\mathbb{R}}^d$. In this basis
\begin{equation}
\Gamma[M,A]= \frac{1}{\beta}\sum_{p_0}\int\frac{d^d\bfp}{(2\pi)^d}
\,\tr\langle p|f(M,D)|p\rangle \,.
\end{equation}

At this point the symbols method can be used (see
e.g. \cite{Salcedo:1996qy,Pletnev:1999yu}): let $|0\rangle$ denote the
state with $p=0$, then using the identities $|p\rangle=
e^{xp}|0\rangle$ (where $e^{xp}$ acts as a multiplicative operator and
the quantities $p_\mu$ are constant c-numbers) as well as
$e^{-xp}D_\mu e^{xp}=D_\mu+p_\mu$, $e^{-xp}M e^{xp}=M$, one obtains
\begin{equation}
\langle p|f(M,D)|p\rangle= 
\langle 0|e^{-xp}f(M,D)e^{xp}|0\rangle= 
\langle 0|f(M,D+p)|0\rangle\,,
\end{equation}
and so the functional trace can be cast in the form
\begin{equation}
\Gamma[M,A]= \frac{1}{\beta}\sum_{p_0}\int\frac{d^d\bfp}{(2\pi)^d}
\,\tr\langle 0|f(M,D+p)|0\rangle \,.
\label{eq:n10}
\end{equation}
In this expression it is clear the requirement of regularity on $f$:
the functional trace comes after integration over momenta and sum over
frequencies and this requires $f$ to be sufficiently convergent for
large $p_\mu$. Let us remark that $|0\rangle$ is periodic rather than
antiperiodic in the temporal direction. The information on whether we
are dealing with bosons or fermions is now contained solely in the
values taken by $p_0$. The state $|0\rangle$ satisfies
\begin{equation}
\langle x|0\rangle = 1\,,\quad
\partial_\mu|0\rangle= \langle 0|\partial_\mu=0\,,\quad
\langle 0|0\rangle = \int d^{d+1}x\,.
\label{eq:o10}
\end{equation}
In addition, when $\hat{h}$ is a multiplicative operator,
$\hat{h}|x\rangle=h(x)|x\rangle$,
\begin{equation}
\langle 0|\hat{h}|0\rangle = \int d^{d+1}x\,h(x)\,.
\label{eq:n13}
\end{equation}
It follows that $\partial_\mu$ appearing inside $f(M,D+p)$ in \eq{n10}
acts derivating everything to its right (or its left, by parts) and
then vanishes after it reaches $|0\rangle$ (or $\langle 0|$). This is
a well-defined working rule and from it stems the usefulness of the
symbols method.

Unfortunately, gauge invariance is no longer manifest when using the
momentum basis. In fact, $\tr\langle 0|f(M,D+p)|0\rangle$ is not gauge
invariant because $|0\rangle$ (or more generally $|p\rangle$) is not
covariant under local transformations. For instance, according to the
rules given in \eq{o10},
\begin{equation}
\tr\langle 0|D_\mu^2|0\rangle= \tr\langle
0|A_\mu^2|0\rangle=\int d^{d+1}x\,\tr[A_\mu^2(x)]
\end{equation}
breaks gauge invariance. However,
\begin{equation}
\tr\langle 0|[D_\mu,D_\nu]^2|0\rangle= \int
d^{d+1}x\,\tr[F_{\mu\nu}^2(x)]
\end{equation}
does not. Note that $[D_\mu,D_\nu]$ is a multiplicative operator
whereas $D_\mu^2$ is not. As a rule, when an operator $g(M,D)$ (a
gauge covariant operator) is multiplicative, \eq{n13} applies and
$\tr\langle 0|g(M,D)|0\rangle$ is gauge
invariant\cite{Salcedo:1996qy}. In \eq{n10} gauge invariance is only
recovered after integration over momenta and sum over
frequencies.\footnote{An elegant method has been presented
in\cite{Pletnev:1999yu} which yields gauge invariant expressions prior
to momentum integration, at the price of introducing derivatives with
respect to $p_\mu$. The method has not yet been extended to include
discrete momenta, as required at finite temperature but in can be
applied to the integration over $\bfp$.}  This will be further
discussed subsequently.

\subsection{The derivative expansion at finite temperature}
\label{subsec:II.C}
By computing the functional trace we essentially mean to end up with
purely multiplicative operators, since this implies that the
functional is expressed as the integral of a function over
space-time. At zero temperature this is usually equivalent to saying
that all derivative operators $D_\mu$ appear inside commutators. In
addition it means to carry out as many implied sums and integrations
(over frequencies and momenta or other parameters) as possible.

In general it is not possible to compute $\Gamma[M,A]$ in closed form
and one must resort to approximations.  The standard approach is to
make power expansions in one or more operators appearing in $f(M,D)$
while the remaining operators are treated non perturbatively. As will
be clear below, a naive expansion in powers of $D_0$ would break gauge
invariance, therefore, because it is in general difficult to work with
two or more non perturbative operators unless they are commuting, and
our present emphasis is in the preservation of manifest gauge
invariance rather than in a particular computation, we will keep $D_0$
as the only operator to be treated non perturbatively, and expand in
all other operators $M$ and $\bfD$.

Before proceeding, let us be more precise about the meaning of
expanding in powers of $M$ and $\bfD$. A convenient way to define the
expansion is by introducing constant c-number bookkeeping parameters,
$f(M,D_0,\bfD)\to f(\lambda_1 M,D_0,\lambda_2\bfD) $, so that counting
powers of those operators is equivalent to counting powers of
$\lambda_{1,2}$. This procedure preserves gauge invariance since it
amounts to a modification of the function $f$. After applying the
symbols method (through $D_\mu\to D_\mu+p_\mu$, cf. \eq{n10}) the
factor $\lambda_2$ will affect $\bfD$ and $\bfp$, however, this can be
brought to the form $f(\lambda_1 M,D_0+p_0,\lambda_2\bfD + \bfp)$ by a
redefinition of $\bfp$. Therefore the expansion can be formulated as
an expansion in powers of $M$ and $\bfD$ in $f(M,D+p)$.\footnote{Of
course, the actual expansion in powers of $\bfD$ must be done after
arriving at \eq{n10} (or other similar formulas in other approaches,
such as Schwinger proper time method) i.e., after $D_\mu$ has become
$D_\mu+p_\mu$, since otherwise powers of $p_\mu$ would be generated as
well and that would destroy the ultraviolet convergence of the
formula.}  Equivalently, the expansion in $\bfD$ can be obtained
directly from the functional $\Gamma[M,A]$ by means of a covariant
spatial dilatation, namely, $M(x_0,\bfx)\to M(x_0,\lambda_2 \bfx)$,
$A_0(x_0,\bfx)\to A_0(x_0,\lambda_2 \bfx)$ and $\bfA(x_0,\bfx)\to
\lambda_2 \bfA(x_0,\lambda_2 \bfx)$. This guarantees that the
expansion is well-defined, i.e., it depends on the functional itself
and not on how it is written or computed.

The situation is completely different for an expansion in powers of
$D_0$. A bookkeeping parameter $D_0\to\lambda_3 D_0$ can be introduced
in $f(M,D)$ and this defines a new $\lambda_3$-dependent gauge
invariant functional. Nevertheless this functional is not useful since
it presents an essential singularity at $\lambda_3=0$, as can be seen
in the simple case of fermions in 0+1 dimensions\cite{Dunne:1997yb}.
(The dependence on $\lambda_{1,2}$ is analytic or at least asymptotic
under suitable regularity conditions on the fields and on the function
$f$.) After applying the symbols method, $f(M,\lambda_3 D_0 +\lambda_3
p_0,\bfD)$ is obtained. However, because $p_0$ is a discrete variable
this is not equivalent to $f(M,\lambda_3 D_0 + p_0,\bfD)$. Therefore
expanding in the explicit $D_0$ in $f(M,D+p)$ does not correspond to a
modification of $f$ and in fact violates gauge invariance.  Also, it
is not possible to introduce $\lambda_3$ by means of a rescaling of
type $x_0\to\lambda_3 x_0$ of the field configuration $(M,A)$ since
this transformation violates the periodicity condition on the
wavefunctions of ${\cal H}$.

After expansion there will be all kind of terms which will be products
of single factors of $M$ and $\bfD$ as well as operators depending non
perturbatively on $D_0$ (by this we merely mean that all orders of
$D_0$ are retained). It is always possible to bring all $\bfD$
operators to the right producing commutators, so that we end up with
two kind of terms: i) terms in which all operators $\bfD$ appear only
in commutators (more precisely, in the form $[\bfD,\ ]$) and ii) terms
with unsaturated factors $\bfD$ at the right (i.e., $\bfD$ not inside
a commutator). The terms of the first type are multiplicative
operators regarding $\bfx$-space, although they are still differential
(or pseudo-differential) operators with respect to $x_0$-space. The
terms of the second type are non-multiplicative in $\bfx$-space. As we
have argued above, these latter terms break gauge invariance and in
fact they will cancel after integration over $\bfp$.  This can be seen
as follows: let us replace $\bfD$ by $\bfD+\bfa$, where $\bfa$ is a
constant c-number. This replacement has no effect on the terms where
$\bfD$ is in commutators, but counts the contribution from the terms
with unsaturated $\bfD$. However, it is clear that there is no such a
contribution after integration over momenta since $\bfa$ can be
compensated by a similar shift in the integration variable
$\bfp$. Thus at the end, all operators $\bfD$ appear in commutators
only. (The same result is obtained directly using the method of
Pletnev and Banin\cite{Pletnev:1999yu}.) A similar argument would
break down for $D_0$: a shift to $D_0+a_0$ cannot, in general, be
compensated by a shift in $p_0$ since at finite temperature the
frequency is a discrete variable.

From the previous discussion it follows that we only have to retain
those terms where all operators $\bfD$ are in commutators. Because all
operators are now multiplicative in $\bfx$-space, $\bfx$ becomes just
a parameter in what follows.  Let us consider a typical term:
\begin{equation}
\hbox{TT}=\frac{1}{\beta}\sum_{p_0}\int\frac{d^d\bfp}{(2\pi)^d}
\,\tr\langle 0|
\alpha_1(D_0+p_0,\bfp)X\alpha_2(D_0+p_0,\bfp)Y\alpha_3(D_0+p_0,\bfp)
|0\rangle  \,.
\label{eq:n14}
\end{equation}
$X$ and $Y$ are multiplicative and gauge covariant operators
constructed with $D_\mu$ and $M$, and the $\alpha_i(x,y)$ are some
functions. At this point the integration over $p_\mu$ is non trivial
(even for the simplest forms of the functions $\alpha_i(x,y)$) because
$p_\mu$ appears in different and non-commuting operators. A possible
approach is to express the operators in terms of their matrix elements
using as basis a complete set of eigenstates of $D_0$ (in the Hilbert
space of time and internal degrees of freedom). These matrix elements
are then ordinary functions of $p_\mu$. Instead of that, we will use
the equivalent prescription of labelling the operators $D_0$ according
to their position with respect to $X$ and $Y$: the symbols $D_{01}$,
$D_{02}$ and $D_{03}$ will be used to denote the operator $D_0$ is
positions 1 (before $X$), 2 (between $X$ and $Y$) and 3 (after $Y$)
respectively. In this notation
\begin{equation}
\hbox{TT}=\frac{1}{\beta}\sum_{p_0}\int\frac{d^d\bfp}{(2\pi)^d}
\,\tr\langle 0|
\alpha_1(D_{01}+p_0,\bfp)\alpha_2(D_{02}+p_0,\bfp)\alpha_3(D_{03}+p_0,\bfp)XY
|0\rangle  \,.
\end{equation}
An immediate consequence is that the labeled operators are effectively
commuting and the momentum integration and frequency summation can be
carried out as for ordinary functions. The result can be written as
\begin{equation}
\hbox{TT} =\tr\langle 0| g(D_{01},D_{02},D_{03})XY |0\rangle
\label{eq:o11}
\end{equation}
where the function $g$ is defined by
\begin{equation}
g(x,y,z) =
\frac{1}{\beta}\sum_{p_0}\int\frac{d^d\bfp}{(2\pi)^d} \,
\alpha_1(x+p_0,\bfp)\alpha_2(y+p_0,\bfp)\alpha_3(z+p_0,\bfp)\,.
\end{equation}
(Note that there will be two versions of $g$, the bosonic one and the
fermionic one, which are related by a shift of $i\pi/\beta$ in their
arguments.) By construction the function $g$ is periodic:
\begin{equation}
g(x,y,z) = 
g(x+\frac{2\pi i}{\beta},y+\frac{2\pi i}{\beta},z+\frac{2\pi i}{\beta})\,.
\label{eq:n20}
\end{equation}
This is an immediate consequence of the sum over Matsubara frequencies
and ultraviolet convergence of the expressions.

This periodicity property is essential to codify the gauge invariance
of the original expression. To see this, let us introduce the
operation $\Dh_\mu$ which is defined as $\Dh_\mu X=[D_\mu,X]$ for any
operator $X$. Consistently with our previous notation, we will denote
by $\Dh_{01}$ the action of $\Dh_0$ in position 1 (i.e., on $X$) and
by $\Dh_{02}$ the action of $\Dh_0$ in position 2 (i.e., on $Y$).
Then clearly
\begin{equation}
\Dh_{01}= D_{01}-D_{02}\,, \quad
\Dh_{02}= D_{02}-D_{03}\,.
\end{equation}
The interesting point is that these formulas hold for arbitrary
functions of $\Dh_{01}$ and $\Dh_{02}$ as well. This follows from the
well-known identity $e^ABe^{-A}=e^{[A,\ ]}B$: for any c-number
$\lambda$
\begin{equation}
e^{\lambda (D_{01}-D_{02})}XY=
e^{\lambda D_0}X e^{-\lambda D_0}Y=
\left(e^{\lambda \Dh_0} X\right) Y=
e^{\lambda \Dh_{01}}XY\,,
\end{equation}
and this identity immediately extends to arbitrary functions of
$D_{01}-D_{02}$, and analogously for $\Dh_{02}$.  This allows to make
everywhere the replacements 
\begin{equation}
D_{02}= D_{01}-\Dh_{01}\,,\quad D_{03}=
D_{01}-\Dh_{01}-\Dh_{02} 
\end{equation}
and use $D_{01}$, $\Dh_{01}$ and $\Dh_{02}$ as the independent
variables to work with. The advantage of doing this is that the action
of $\Dh_{01}$ and $\Dh_{02}$ on $X$ and $Y$ produces multiplicative
and gauge covariant operators. On the other hand, the presence of the
operator $D_{01}$, which is outside commutators, combined with the
gauge non-covariant operation $\langle 0|\ |0\rangle$, still can
introduce gauge non-invariant contributions. This is avoided thanks to
the periodicity property of $g(x,y,z)$ as we will show now. Indeed,
the periodicity property allows to write the term as
\begin{equation}
\hbox{TT} 
=\tr\langle 0| \varphi(e^{-\beta D_{01}},\Dh_{01},\Dh_{02})XY
|0\rangle \,,
\label{eq:n24} 
\end{equation}
where the function $\varphi$ is defined by
\begin{equation}
\varphi(e^{-\beta x},y, z)= g(x,x-y,x-y-z) \,.
\end{equation}
The periodicity condition of $g$ ensures that the function
$\varphi(\omega,y,z)$ is one-valued (it depends on $\omega$ and not
just on $\log(\omega)$). In order to bring the expression into a
manifestly gauge invariant form, we will use the following property:
\begin{equation}
e^{-tD_0}= e^{-t\partial_0}\,
Te^{-\int_{x_0}^{x_0+t}A_0(x_0^\prime,\bfx)dx_0^\prime}\,,\quad t\ge 0
\,.
\end{equation}
Here $t$ is just a parameter and $T$ denotes time ordered
product. (The quantities $D_0$, $\partial_0$ and $x_\mu$ represent
operators and the product refers to a product of operators so that
$\partial_0$ is not directly derivating $x_0$.)  This equation can be
easily proved by the standard procedure of showing that the two
expressions satisfy the same first order differential equation in $t$
and coincide at $t=0$. The left-hand side is a manifestly gauge
covariant operator. It is interesting to see how gauge covariance is
realized in the right-hand side: the time-ordered product from $x_0$
to $x_0+t$ (with $\bfx$ fixed) transforms with $U(x_0,\bfx)$ at the
right and $U^{-1}(x_0+t,\bfx)$ at the left, and this latter factor is
transformed into $U^{-1}(x_0,\bfx)$ after commutation with
$e^{-t\partial_0}$, thus the product of the two factors transforms
covariantly at $(x_0,\bfx)$, as $e^{-t D_0}$.

In particular, by taking $t=\beta$ in the previous formula, one
obtains the identity
\begin{equation}
e^{-\beta D_0}= e^{-\beta \partial_0}\,\Omega \,,
\end{equation}
where
\begin{equation}
\Omega(x)=
T\exp\left(-\int_{x_0}^{x_0+\beta}A_0(x_0^\prime,\bfx)dx_0^\prime\right).
\label{eq:28}
\end{equation}
(Again $e^{-\beta \partial_0}\,\Omega$ is to be understood as the
product of two operators.) Beyond the interval $[0,\beta]$ $A_0(x)$ is
defined as a periodic function of the time, so $\Omega(x)$ is also
periodic. Although $\Omega(x)$ is non-local in terms of $A_0$, it
behaves as a local field which takes values on the gauge group. In
particular it transforms covariantly at $x$:
\begin{equation}
\Omega^U(x)=U^{-1}(x)\Omega(x)U(x)\,.
\end{equation}
The matrices $\Omega(x)$ at different values of $x_0$, but equal
$\bfx$, are related by similarity transformations and their trace, the
Polyakov loop, is independent of $x_0$. Another important property is
\begin{equation}
\Dh_0\Omega= [D_0,\Omega]= 0\,.
\label{eq:n28}
\end{equation}

On the other hand, the effect of the operator $\exp(-\beta
\partial_0)$ is to produce the shift $x_0\to x_0-\beta$, therefore it
is equivalent to the identity operator on the space of periodic
functions in which we are working (as noted the periodic wavefunction
$|0\rangle$ appears regardless of whether we are considering bosons of
fermions). So in this space
\begin{equation}
e^{-\beta D_0}= \Omega \,.
\end{equation}
This produces the manifestly gauge invariant expression
\begin{equation}
\hbox{TT} 
=\tr\langle 0|\varphi(\Omega_1,\Dh_{01},\Dh_{02})XY
|0\rangle \,.
\end{equation}
(The label 1 in $\Omega$ indicates to put this operator in position
1. The relative order between $\Dh_{01}$ and $\Omega_1$ is immaterial
due to \eq{n28}.)

It can be noted that all previous manipulations, starting from
\eq{n14}, hold also without taking $\tr\langle 0|\, |0\rangle$. Inside
$\tr\langle 0|\, |0\rangle$ integration by parts implies that
$\Dh_{01}$ is equivalent to $-\Dh_{02}$ (or equivalently, that
$D_{03}=D_{01}$), and so we have the final formula
\begin{equation}
\hbox{TT} 
=\tr\langle 0|\varphi(\Omega_1,\Dh_{02})XY
|0\rangle \,,
\label{eq:n33}
\end{equation}
where
\begin{equation}
\varphi(\omega,y)= \varphi(\omega, -y, y)\,.
\end{equation}
The whole point of these manipulations was to end up with a manifestly
gauge covariant and multiplicative operator so that
$\tr[\varphi(\Omega_1,\Dh_{02})XY]$ is just a gauge invariant function
of $x$, constructed with the fields $A_\mu(x)$ and $M(x)$:
\begin{equation}
\hbox{TT} =\int d^{d+1}x\,\tr[\varphi(\Omega_1,\Dh_{02})XY] \,,
\label{eq:n35}
\end{equation}

The fact that $g(x,y,z)$ is periodic is essential to produce the gauge
covariant operator $\Omega(x)$. This periodicity property would be
lost in an expansion in powers of $D_0$. We should remark that no
restriction has been put on the field configuration, which is
completely general (may be non Abelian and non stationary) and also no
choice of gauge has been needed.

Perhaps it should be emphasized how exactly the lack of periodicity
would break gauge invariance. To see this it is sufficient to consider
$0+1$-dimensional fermions in presence of an Abelian configuration
\cite{Dunne:1997yb}. The corresponding effective action is of the form
$\beta g(a)$ with $a=\int_0^\beta dx_0 A_0$. The quantity $a$ is
invariant under topologically small gauge transformations, $A^U_0 =A_0
+\partial_0\Lambda$ ($\Lambda(x_0)$ being a periodic function) but
under a large gauge transformation, e.g. $A^U_0 =A_0 + 2\pi i n/\beta$
(which corresponds to $U(x_0)=\exp(2\pi i nx_0/\beta)$), $a$ changes
by an integer multiple of $2\pi i$, so $g(a)$ will not be invariant in
general. When $g$ is periodic the effective action becomes
$\beta\varphi(\Omega)$ with $\Omega=\exp(-a)$ and it is invariant
under all gauge transformations. See Section \ref{subsec:IV.A} for
further remarks.

\subsection{Relation with the calculation fixing the gauge}

In \cite{Salcedo:1999sv,Salcedo:1998tg} the kind of calculation just
described was carried out for fermions but fixing the gauge through
the gauge condition $\partial_0A_0=0$. (The idea was that the two
operators treated not perturbatively, $\partial_0$ and $A_0$, are then
commuting.) No loss of generality is actually implied by this approach
since such a gauge always exists\cite{Salcedo:1999sv}. However,
because it is not unique, it is necessary to find all remaining gauge
transformations allowed within the $A_0$-stationary gauge, and then
check that all of them produce the same result. This was shown to be
equivalent to the periodicity condition that follows from summing over
Matsubara frequencies, \eq{n20}. All this is unnecessary in the
present approach since the gauge has not been fixed. Using gauge
invariance, the results obtained within the $A_0$-stationary gauge can
directly be taken over as follows. When $\partial_0A_0=0$, the field
$\Omega$ becomes $e^{-\beta A_0}$, so it is only necessary to replace
$e^{-\beta A_0}$ of the calculation in $A_0$-stationary gauge by
$\Omega(x)$ to obtain the result expressed in an arbitrary gauge.
Further comments are made in \ref{subsec:IV.A}.

\subsection{Expansion in space-time derivatives at finite temperature}

As noted, expanding in powers of $D_0$ breaks periodicity and hence
gauge invariance, however, in principle nothing prevents from
expanding in powers of $\Dh_0$ in \eq{n35}, namely,
\begin{equation}
\hbox{TT} 
= \sum_{n=0}^\infty\int d^{d+1}x\,\tr[\varphi_n(\Omega)X\Dh_0^nY] \,,
\label{eq:n36}
\end{equation}
where
\begin{equation}
\varphi(\omega,y)=\sum_{n=0}^\infty \varphi_n(\omega)y^n \,.
\end{equation}
The interest of doing that is, of course, that, at least at lower
orders, the result is simpler than the full result. This can be
regarded as the finite temperature generalization of the usual
derivative expansion at zero temperature. Recall that $\bfD$ already
was restricted to appear in commutators only, so this is really an
expansion in powers of $\Dh_\mu$. As usual, higher orders are
increasingly ultraviolet convergent.  It can be noted that in the
Abelian and stationary case $\Dh_0$ vanish identically (on
multiplicative operators, such as $X$ and $Y$ in \eq{n36}) therefore
the zeroth order in the above expansion becomes exact.

Nevertheless, it should be noted that such an expansion is not as
well-defined as for instance the expansion in powers of $\bfD$. The
latter is defined from the functional itself, since it corresponds to
spatial dilatations of the fields.  No such transformation is known
for the expansion in powers of $\Dh_0$. So it principle it should be
expected that different ways of expressing the functional in terms of
$\Omega$ and $\Dh_0$ would yield different expansions, only the sum of
all orders being unambiguously defined. This can be seen more clearly
as follows. Recalling that inside $\langle 0| |0\rangle$ the operators
$D_{03}$ and $D_{01}$ are equivalent, the typical term considered
above, \eq{o11}, takes the form
\begin{equation}
\hbox{TT} =\tr\langle 0| g(D_{01},D_{02})XY |0\rangle
\end{equation}
(with $g(x,y)= g(x,y,x)$).
Using the final form \eq{n33}, it is easily established that it can
also be written as
\begin{equation}
\hbox{TT} =\tr\langle 0| g(D_{02},D_{01})YX |0\rangle \,,
\end{equation}
because in \eq{n33} all operators are multiplicative and therefore
integration by parts and the cyclic symmetry can be used. Now, in the
frequent case of contributions where $X=Y$, this implies that only the
symmetric part of the function $g(x,y)$ is actually
contributing. However, it is easy to write purely antisymmetric and
periodic functions $g(x,y)$ such that when used in \eq{n36}, each
order is non vanishing, although of course their full contribution
vanish when summed to all orders.  This particular kind of ambiguity
can be fixed by imposing a symmetry restriction on $g(x,y)$ before
carrying out the expansion in $\Dh_0$. This ambiguity is further
discussed in Section \ref{subsec:IV.B}

\subsection{Illustration of the method}
\label{subsec:II.F}

To illustrate the previous manipulations in a practical case, we will
consider the quantity
\begin{equation}
C[m,A]= -\frac{1}{4}\Tr_b\left[\left(
 \frac{1}{-D_\mu^2+m^2}
\frac{1}{2}\sigma_{\mu\nu}F_{\mu\nu}
\right)^2\right] + O(F_{\mu\nu}^3) \,,
\label{eq:5}
\end{equation}
which will appear later in the study of the scalar field in 2+1
dimensions. In this expression $F_{\mu\nu}=[D_\mu,D_\nu]$ and
$\sigma_{\mu\nu}=\frac{1}{2}[\gamma_\mu,\gamma_\nu]$ (where
$\gamma_\mu$ are Hermitian Dirac matrices in 2+1 dimensions).
$m$ is a c-number.

$C[m,A]$ is ultraviolet finite and so it is a well-defined and
unambiguous quantity. We will compute it through second order in an
expansion in the number of spatial covariant derivatives. Clearly the
expansion starts at second order and it is sufficient to retain the
explicit term in \eq{5} since terms of $O(F_{\mu\nu}^3)$ must contain
four or more spatial indices.

First, the symbols method, \eq{n10}, is applied. Afterwards, taking
the trace in Dirac space and keeping just terms with two spatial
covariant derivatives, produces
\begin{equation}
C_2[m,A]= \frac{1}{2}
\int\frac{ d^2\bfp}{(2\pi)^2}\frac{1}{\beta}\sum_{p_0} 
\langle 0| \tr\left[\left(
 \frac{1}{-(D_0+p_0)^2+\bfp^2+m^2} \bfE
\right)^2\right] |0\rangle \,,
\end{equation}
where $\bfp^2=-p_i^2$, $p_0= 2\pi i n/\beta$, and
$\bfE =[D_0,\bfD]$.

To proceed to the integration over momenta and sum over frequencies we
use the trick of adding a label 1 or 2 to the operators $D_0$ to
indicate their actual position in the expression, namely
\begin{eqnarray}
C_2[m,A] &=& \frac{1}{2} 
\int\frac{ d^2\bfp}{(2\pi)^2}\frac{1}{\beta}\sum_{p_0} 
\langle 0|\tr\left[
 \frac{1}{-(D_{01}+p_0)^2+\bfp^2+m^2} 
\frac{1}{-(D_{02}+p_0)^2+\bfp^2+m^2} \bfE^2
\right] |0\rangle
\nonumber\\
&:=& 
\langle 0|\tr\left[g(D_{01},D_{02}) \bfE^2
\right] |0\rangle
\end{eqnarray}

The momentum integration yields
\begin{equation}
g(x_1,x_2)= -\frac{1}{8\pi}
\frac{1}{\beta}\sum_{p_0} 
 \frac{1}{(x_1+p_0)^2-(x_2+p_0)^2} 
\log\left(\frac{m^2-(x_1+p_0)^2}{m^2-(x_2+p_0)^2}\right) \,.
\label{eq:10}
\end{equation}
In order to sum over frequencies, it is convenient to reduce the
expression to a rational form. This is achieved by derivating with
respect $m$ and then integrating back (using that $g(x_1,x_2)$
vanishes as $m\to\infty$). Then the identity
\begin{equation}
\sum_n\left(\frac{1}{x_1+i\pi n}-\frac{1}{x_2+i\pi n}\right)
=\coth(x_1)-\coth(x_2)\,,
\end{equation}
can be applied. This produces
\begin{equation}
g(x_1,x_2) =
-\frac{1}{16\pi} \frac{1}{x_1-x_2}
\int_m^\infty dt 
\Bigg(
\frac{
\coth(\frac{\beta}{2}(t+x_1))
}{2t+x_1-x_2}
-\frac{\coth(\frac{\beta}{2}(t+x_2))
}{2t-x_1+x_2}
\Bigg)
+\hbox{p.p.c.}
\label{eq:6}
\end{equation}
Where p.p.c. (which stands for pseudo-parity conjugate) refers to the
same expression with the replacements $x_1\to -x_1$ and $x_2\to -x_2$.

Correspondingly,
\begin{equation}
C_2[m,A]= \int d^3x \,\tr\left[\varphi(\Omega_1,\Dh_{02}) \bfE^2
\right]
\label{eq:n43}
\end{equation}
with
\begin{equation}
\varphi(\omega,y)=
\frac{1}{16\pi} \frac{1}{y}
\int_m^\infty dt 
\Bigg(
\frac{1}{2t-y}\frac{e^{\beta t}+\omega}{e^{\beta t}-\omega}
-\frac{1}{2t+y}\frac{e^{\beta (t+y)}+\omega}{e^{\beta(t+y)}-\omega}
\Bigg)
+\hbox{p.p.c.}
\label{eq:n6}
\end{equation}
p.p.c. corresponds to $y\to -y$ and $\omega\to \omega^{-1}$.  This is
the final expression which contains all contributions to $C[m,A]$ with
two spatial Lorentz indices and any number of zeroth indices.  As
expected at finite temperature, it is non-local in time but local in
$\bfx$. Note that $C_2[m,A]$ is an even function of $m$.\footnote{This
can be shown by noting that $\varphi(\omega,y;m)-\varphi(\omega,y;-m)$
is given by the same formula (\ref{eq:n6}) with replacement
$m\to-\infty$, and then showing that the integrand is convergent and
odd as a function of $t$.}

We can now consider a further expansion in powers of $\Dh_0$ since it
respects gauge invariance. An explicit computation shows that
$\varphi(\omega,y)$ is an analytic function of $y$. At leading
(zeroth) order in $\Dh_0$ the result from \eq{n6} is
\begin{equation}
C_2[m,A]= 
\frac{1}{16\pi}\frac{1}{2m}
\int d^3x \,\tr\left[\left(
\frac{e^{\beta m}+\Omega}{e^{\beta m}-\Omega}
+\frac{e^{\beta m}+\Omega^{-1}}{e^{\beta m}-\Omega^{-1}}
\right) \bfE^2 \right] 
 +O(\Dh_0)
\,.
\label{eq:11}
\end{equation}
Note that this result, unlike the full result in
eqs. (\ref{eq:n43},\ref{eq:n6}), does not contain an integral over the
mass ($\int_m^\infty dt$). This property holds to all orders in
$\Dh_0$.

\subsection{The effective action}

In this subsection we summarize some properties of the effective
action which will be needed later.

{\sl The effective action.} The Euclidean effective action is defined
as minus the logarithm of the partition functional. For non
interacting fields it takes the form
\begin{equation}
W[M,A]= c\,\Tr\log(K(M,D))=c\,\log\,\Det(K(M,D))\,,
\label{eq:n47}
\end{equation}
where $K(M,D)$ is a differential operator (e.g. the Klein-Gordon
operator as in \eq{n2} or the Dirac operator) and $c$ some
constant. 

{\sl Ultraviolet ambiguities.} The previous expression needs to be
regularized, and a number of methods can be used to obtain a
renormalized version of it. The key observation is that all
renormalized versions of the effective action must yield the same
ultraviolet finite contributions and so two such versions can differ
at most by a term which is a local polynomial in the external fields
and their derivatives, with canonical dimension no more than $d+1$ (in
$d+1$ dimensions). Note that for this discussion a mass $m$ plays the
role of an external scalar field which happens to take a constant
c-number configuration and so, in particular, the ambiguity in the
renormalized action will depend polynomically on $m$. This implies
that any sensible (that is, correctly describing the ultraviolet
finite contributions) regularization plus renormalization prescription
can be used to make the effective action finite; the actual effective
action describing the physical system at hand will correspond to
adding the appropriate local polynomial action to the previous
result. Another consequence is that formal identities can be applied
so long as a violation of them is allowed in the form of a local
polynomial of dimension $d+1$ or less.

{\sl The $\zeta$-function method.} The effective action can be defined
through the $\zeta$-function prescription, namely
\begin{equation}
\Tr\log(K) = \frac{d}{ds}\Tr(K^s)\Big|_{s=0}\,.
\label{eq:new.10a}
\end{equation}
(An analytical extension in $s$ is understood from sufficiently
negative values of $s$.) When $K$ admits a complete set of
eigenvectors, $\Tr(K^s)=\sum_n\lambda_n^s$, $\lambda_n$ being the
eigenvalues of $K$. If the calculation has to be made using some
expansion it is convenient to use the following formula
\cite{Seeley:1967ea}
\begin{equation}
\Tr(K^s) = \Tr\int_\Gamma\frac{dz}{2\pi i}z^s\frac{1}{z-K}\,,
\label{eq:new.10}
\end{equation}
where the path $\Gamma$ encloses anti-clockwise the eigenvalues of $K$
but not $z=0$. This method is practical in actual calculations
combined with the symbols method: after applying \eq{n10} to expand
the functional trace, it is straightforward to make an explicit
expansion in $M$ and $\bfD$, for instance. This method has been used
for fermions in \cite{Salcedo:1996qy} and for a non-local Dirac
operator in \cite{RuizArriola:1999zi} at zero temperature, and at
finite temperature for odd dimensional fermions in
\cite{Salcedo:1999sv} and even dimensional fermions in
\cite{Salcedo:1998tg}.

{\sl Anomalies.} If the effective action breaks a symmetry of the
action there is an anomaly. In general the anomaly can be defined as
the difference between the effective action of the original and the
transformed configurations (of the external fields) and by
construction is a local polynomial. It may happen that the symmetry
can be restored by adding an appropriate local polynomial to the
effective action. In this case the breaking is an unessential
anomaly. When not all symmetries can be restored simultaneously the
theory presents an essential anomaly. All symmetries which are
implemented through a similarity transformation of $K(M,D)$ leave the
spectrum invariant, thus, because the $\zeta$-function prescription
defines the determinant of a differential operator by regularizing the
product of its eigenvalues, it follows that for these symmetries the
$\zeta$-function version of $\Det(K)$ is always free of anomalies.
This applies in particular to vector gauge invariance. (Axial gauge
transformations and scale transformations, for instance, are not
implemented by similarity transformations and they can be anomalous.)
Therefore the partition functional is always invariant under gauge
transformations. On the other hand the effective action can change by
integer multiples of $2\pi i$, since the logarithm is a many-valued
function. (We are assuming that no zero modes are involved. They would
induce changes multiples of $i\pi$ in the effective action.) By
continuity, this can only happen for topologically large gauge
transformations. For a scalar field the multivaluation cannot occur
since the corresponding $\zeta$-function renormalized effective action
is purely real. For fermions the multivaluation may take place
depending on the topological numbers of the gauge transformation and
the gauge field configuration and this indicates the presence of
topological pieces in the effective action
~\cite{Deser:1982vy,Redlich:1984kn,Niemi:1983rq}. Such multivaluation
is indeed found in the explicit calculation for $2+1$-dimensional
fermions of \cite{Salcedo:1999sv}.

{\sl Locality and finite temperature.} In previous subsections we have
considered operators of the form $f(M,D)$ with $f$ one-valued and
ultraviolet convergent.  Actually, one wants to compute the effective
action which contains multivaluation and ultraviolet divergences,
\eq{n47}. In many expansions (perturbative, derivative, $1/m$, etc)
higher orders are ultraviolet finite and thus they are also free of
multivaluation. For those terms all our previous considerations hold
directly. In particular, we find a remarkable result, namely, that the
effective action at finite temperature can be written as
(cf. \eq{n36})
\begin{equation}
W[M,A] = \sum_n\int d^{d+1}x\,\tr[\varphi_n(\Omega){\cal O}_n] \,,
\label{eq:n36a}
\end{equation}
where $\varphi_n$ are some functions and ${\cal O}_n$ are gauge
covariant and local operators constructed out of $\Dh_\mu$ and $M$. In
this sense the theory at finite temperature is local in the usual
sense (i.e., the effective action admits an expansion in $\Dh_\mu$)
provided that the field $\Omega(x)$ is regarded as local.

For the ultraviolet divergent terms, some oddities appear which are
necessary in order to accommodate the existence of anomalies,
topological terms and multivaluation, all these issues being
related. For instance, when the expansion in spatial covariant
derivatives is computed for fermions in $2+1$
dimensions\cite{Salcedo:1999sv} the functions $\varphi_n$ of lower
orders are many-valued (a property belonging to the exact result in
$0+1$ dimensions \cite{Dunne:1997yb} and in $2+1$ dimension for
Abelian and stationary configurations \cite{Fosco:1997ei}). In
addition, negative powers of $\Dh_0$ may appear when one goes beyond
the Abelian and stationary case. This was handled in
\cite{Salcedo:1999sv} by introducing the fields $\bfAr(x)$, defined as
any solution of the equation $\Dh_0 \bfE=\Dh_0^2\bfAr$. There it was
shown that the ambiguity in the definition of $\bfAr$ cancels and all
solutions yield the same effective action. In terms of the $\bfAr(x)$
and $\Omega(x)$ the finite temperature effective action remains local.

\section{The scalar field in 2+1 dimensions}
\label{sec:3a}

In what follows we will apply the previous ideas to the computation of
the effective action of $2+1$-dimensional scalar particles at finite
temperature and density. The Euclidean action of the system is that of
\eq{n1}.  There, the field $\phi(x)$ is a Lorentz scalar and a vector
in the internal symmetry space which will collectively be referred to
as flavor space, with dimension $N_f$. The covariant derivative is
defined as $D_\mu=\partial_\mu+A_\mu$, where the gauge field
$A_\mu(x)$ is an antihermitian matrix in flavor
space. Correspondingly, the gauge transformations $U(x)$ are unitary
matrices. The mass $m$ is a space-time constant and a real c-number in
flavor space. The more general case of an arbitrary scalar field
$M(x)$ replacing $m$ will not be considered here. The effective action
is given by \eq{n2}.

Relevant symmetries of the problem are pseudo-parity and gauge
transformations. Pseudo-parity corresponds to changing every Lorentz
zeroth index, i.e, $(x_0,\bfx)\to (-x_0,\bfx)$ and $A_0\to
-A_0$. Since the spectrum of the Klein-Gordon operator $-D_\mu^2+m^2$
is unchanged under this transformation the $\zeta$-function
regularization prescription provides a pseudo-parity preserving
effective action. Such Euclidean effective action contains only
contributions with an even number of Lorentz zeroth indices, it does
not contain the Levi-Civita pseudo-tensor, and thus it is purely
real. Any other renormalization prescription can only produce
imaginary contributions which are local polynomials. As noted, the
$\zeta$-function regularized effective action will be strictly gauge
invariant since it is real. In fact no essential anomalies are present
in the case of scalar fields in 2+1 dimensions (scale anomalies are
absent in odd dimensions~\cite{Blau:1988kv}).

There is a third symmetry, namely, the transformation $m\to -m$ which
is trivial for scalar particles and again free from anomaly using
$\zeta$-function regularization. Within other renormalization schemes
there can appear terms breaking this symmetry but they will be
removable by adding a local polynomial. In the case of odd-dimensional
fermions neither pseudo-parity nor the transformation $m\to -m$ are
symmetries, however their product gives the parity
transformation. Parity is a symmetry of the fermionic action but is
not a similarity transformation of the Dirac operator, so it is not
guaranteed to be preserved by the $\zeta$-function renormalization
prescription. As is well-known, parity for odd dimensional fermions is
in general in conflict with invariance under large gauge
transformations and if the latter invariance is enforced, parity may
present an anomaly, depending on the number of
flavors~\cite{Redlich:1984kn}.

\subsection{The 0+1 dimensional model}
\label{sec:2}

The above-mentioned remarks can be illustrated with the 0+1
dimensional version of the system. The corresponding effective action
has been computed in ~\cite{Barcelos-Neto:1999zv}. Perhaps the
simplest way to derive this effective action is by computing the
partition function of the associated 0-dimensional Hamiltonian system
in a gauge where $A_0$ is time independent. The energy spectrum is
then obtained directly from the Klein-Gordon equation written as
\begin{equation}
-\partial_0\phi=(\pm m+A_0)\phi \,.
\end{equation}
Thus for each flavor there are two single particle levels
$\epsilon_{\pm,a}= \pm m+ A_{0,a}$ ($a=1,\dots,N_f$ labelling the
eigenvalues of the matrix $A_0$ in flavor space). The standard textbook
result for the partition function of a system of non-interacting
bosons then applies
\begin{equation}
Z_s[m,A]= \prod_{\sigma=\pm}\prod_{a=1}^{N_f}\sum_{n=0}^\infty
e^{-\beta(\sigma m+A_{0,a})n} \,,
\end{equation}
or equivalently
\begin{equation}
W_s[m,A]=
\tr\,\log\left[\left(1-e^{-\beta(m+A_0)}\right)
\left(1-e^{-\beta(-m+A_0)}\right)\right]\,.
\label{eq:2}
\end{equation}
The trace refers to flavor space. This result can be
rewritten as
\begin{equation}
W_s[m,A]= -\beta\tr(A_0)+\Gamma_s[m,A]\,.
\label{eq:1}
\end{equation}
The first term is the 0+1 dimensional Chern-Simons action which breaks
pseudoparity and can be removed by a local polynomial counterterm. The
second term is (up to a constant)
\begin{equation}
\Gamma_s[m,A]=
\tr\,\log\left[4\sinh\left(\frac{\beta}{2}(m+A_0)\right)
\sinh\left(\frac{\beta}{2}(m-A_0)\right)\right]\,.
\end{equation}
This effective action is an even function of $m$ and $A_0$, so it
preserves parity and pseudo-parity. It can be written in a manifestly
gauge invariant form as
\begin{equation}
\Gamma_s[m,A]=
 \tr\,\log\left[e^{\beta m}+e^{-\beta m} -\Omega-\Omega^{-1}\right]\,.
\label{eq:1b}
\end{equation}
As in the case of fermions\cite{Dunne:1997yb}, periodicity of the
effective action as a function of $\beta A_0$, would be lost within a
perturbative expansion, i.e. an expansion in powers of $A_0$.

Since $\Gamma_s[m,A]$ enjoys all symmetries of the action it coincides
with the $\zeta$-function regularized effective action, up to a
constant (since any other local polynomial must be of degree one in
$m$ or $A_0$ and would break parity). The result in
\cite{Barcelos-Neto:1999zv} corresponds to
$\Gamma_s[m,A]-\Gamma_s[m,0]$ in Minkowski space. It is noteworthy
that the partition function defined directly from the Hamiltonian
breaks pseudo-parity (due to the Chern-Simons term in \eq{1}) even if no
ultraviolet divergences are introduced in 0+1 dimensions in the
canonical formalism.

\subsection{Computation of the effective action in 2+1 dimensions.
Relation to the fermionic case}
\label{sec:3}

The effective action of the 2+1 dimensional model cannot be computed
in closed form for arbitrary space-time and internal symmetry
configurations. Our approach will be to  expand the effective action
in the number of spatial covariant derivatives, or equivalently, in
the number of spatial Lorentz indices. The computation will be carried
out through second order, that is, we will keep terms with zero or two
Lorentz indices. (There are no odd order terms in the expansion.) The
zeroth Lorentz index dependency is treated exactly since this
guarantees the preservation of the periodicity condition which is
essential for gauge invariance.

The calculation can be done by applying the $\zeta$-function
regularization prescription with the help of the symbols method, as
described in ~\cite{Salcedo:1999sv}, however, it is more economical to
use the results already established for fermions in that
reference. This can be done as follows. The effective action for
fermions is
\begin{equation}
W_f[m,A]= -\Tr_f\log(\gamma_\mu D_\mu+m)\,.
\end{equation}
The functional trace is taken in the space of antiperiodic wave
functions and includes Dirac degrees of freedom, in addition to
space-time and flavor degrees of freedom. The gamma matrices are
hermitian and satisfy $\gamma_\mu\gamma_\nu= \delta_{\mu\nu}
+\sigma_{\mu\nu}$. Actually, there are two inequivalent irreducible
representations of the Dirac algebra which are distinguished by the
label $\eta=\pm 1$ in the relation $\gamma_\mu\gamma_\nu\gamma_\rho=
i\eta\epsilon_{\mu\nu\rho}$. So if $\gamma_\mu$ is one of the
representations, $-\gamma_\mu$ provides another inequivalent
representation of the Dirac algebra. The label $\eta$ is attached to
the Levi-Civita pseudo-tensor and thus a change in $\eta$ is
equivalent to a  pseudo-parity transformation. Therefore the fermionic
effective action can be split into two components
\begin{equation}
W_f[m,A]= W_f^+[m,A]+\eta W_f^-[m,A]\,,
\label{eq:4}
\end{equation}
where $W_f^+$ is real and even under pseudo-parity and $W_f^-$ is
imaginary and pseudo-parity odd. (Of course, this relation can be
violated by adding a local polynomial.) Next, note that the formal
identity $\Tr \,\log(AB)= \Tr \,\log(A) +\Tr \,\log(B)$ holds for the
functional trace up to ultraviolet divergent contributions and so it
holds modulo local polynomial terms. This implies that

\begin{eqnarray}
W_f^+[m,A] &=&
-\frac{1}{2}\Tr_f\log\left[ (\gamma_\mu D_\mu+m)(-\gamma_\mu D_\mu+m)
\right]
\nonumber \\  &=&
-\frac{1}{2}\Tr_f\log\left[
-D_\mu^2+m^2-\frac{1}{2}\sigma_{\mu\nu}F_{\mu\nu}
\right]
 \,,  \qquad(F_{\mu\nu}=
[D_\mu,D_\nu])\,.
\label{eq:3}
\end{eqnarray}
So $W_s[m,A]$ (cf. \eq{n2}) is closely related to $W_f^+[m,A]$. The
differences between both expressions are i) the Dirac degree of
freedom which is absent in the scalar case, ii) the different
(periodic versus antiperiodic) boundary conditions and iii) the extra
term $-\frac{1}{2}\sigma_{\mu\nu}F_{\mu\nu}$ which is not present in
the Klein-Gordon operator. In addition, a local polynomial action can
be further added at the end.\footnote{Alternatively, one can choose to
change $m\to -m$ instead of $\gamma_\mu\to -\gamma_\mu$ in the second
factor in the logarithm in \eq{3}, and then relate $W_s[m,A]$ to
$W_f[m,A]+W_f[-m,A]$. Up to a local polynomial, this procedure is
equivalent to the one used in the text.}

The first correction amounts to dividing the fermionic result by the
trace of unity in Dirac space which is two in the 2+1 dimensional
calculation in ~\cite{Salcedo:1999sv}. The second correction can also
be tackled straightforwardly. The Matsubara frequency (related to
$\partial_0$ in $D_0$) takes discrete values $2\pi
i(n+\frac{1}{2})/\beta$ for fermions and $2\pi in/\beta$ for
bosons, thus the functional trace computed in the fermionic Hilbert
space is related to the bosonic one after the replacement $A_0\to A_0
-\frac{i\pi}{\beta}$. This is equivalent to $\Omega(x)\to
-\Omega(x)$. ($\Omega(x)$ has been defined in \eq{28}.)
\begin{equation}
W_s[m,A]= -W_f^+[m,A_0-\frac{i\pi}{\beta},\bfA]-C[m,A]\,.
\label{eq:12}
\end{equation}
The term $C[m,A]$ takes into account the spurious contributions coming
from $-\frac{1}{2}\sigma_{\mu\nu}F_{\mu\nu}$, which have to be removed
from the fermionic result.

This formula can be illustrated in the 0+1 dimensional model, where it
reads $W_s[m,A]= -2W_f^+[m,A_0-\frac{i\pi}{\beta}]$ (note that Dirac
space is 1-dimensional in 0+1 dimensions so the factor 2 is not
canceled in this case, and also $C[m,A]=0$). The simplest way to
obtain the 0+1 dimensional fermionic effective action is again using
the Hamiltonian formalism (fixing $\partial_0A_0=0$). Since there is a
single-particle level with energy $\epsilon_a=\eta m+A_{0,a}$ for each
flavor (where $\eta=\pm 1$ is the Dirac matrix $\gamma_0$ in 0+1
dimensions) it follows that
\begin{equation}
W_f[m,A]= 
-\log
\prod_{a=1}^{N_f}\sum_{n=0,1}
e^{-\beta(\eta m+A_{0,a})n}= 
-\tr\,\log\left[1+e^{-\beta(\eta m+A_0)}\right]\,.
\end{equation}
The result in \cite{Dunne:1997yb} corresponds (up to a local
polynomial) to $W_f[m,A]-W_f[m,0]$ with $\eta=1$.  This version of the
effective action does not directly satisfy \eq{4}, i.e. the
pseudo-parity transformation $A_0\to-A_0$ is not equivalent to the
transformation $\eta\to -\eta$ in the previous formula. However,
subtracting an appropriate $\eta$-dependent polynomial\footnote{To
wit, $\theta(-\eta)\int dx_0\,\tr(\eta m+A_0)$, which is temperature
independent. This is consistent with the fact that the finite
temperature does not introduce new ultraviolet divergences.} yields
\begin{equation}
W_f^\prime[m,A]= -\tr\,\log\left[1+e^{-\beta(m+\eta A_0)}\right]
=-\tr\,\log\left[1+e^{-\beta m}\Omega^\eta \right]\,,
\label{eq:o1}
\end{equation}
which does satisfy \eq{4}. This is the $\zeta$-function result
\cite{Salcedo:1999sv}. It is readily verified that $\Gamma_s[m,A]$ in
\eq{1b} coincides with $-2W_f^{\prime +}[m,A_0-\frac{i\pi}{\beta}]$
plus a polynomial, $\int dx_0\tr(m)$.

In 2+1 dimensions the subtraction $C[m,A]$ can be computed in an
expansion in powers of $\frac{1}{2}\sigma_{\mu\nu}F_{\mu\nu}$.  The
term of first order does not contribute (since the trace of
$\sigma_{\mu\nu}$ in Dirac space vanishes) thus the leading term is
that with two powers of $F_{\mu\nu}$, namely, the expression in
\eq{5}. Because $C[m,A]$ is ultraviolet finite it is free from
anomalies, i.e., it enjoys all symmetries of the bosonic action.  This
term has been computed, up to two spatial derivatives, in subsection
\ref{subsec:II.F}.

Let us quote the results for the pseudo-parity even component of the
effective action of fermions in 2+1 dimensions
\cite{Salcedo:1999sv}. At zeroth order in the number of spatial
covariant derivatives the result is
\begin{equation}
W_{f,0}[m,A] = \frac{1}{4\pi}\tr\langle 0|\left[
\left(\frac{2}{\beta}\right)^2m\phi_1(\frac{\beta}{2}(m-D_0))-
\left(\frac{2}{\beta}\right)^3\phi_2(\frac{\beta}{2}(m-D_0)) \right]
|0\rangle +{\rm p.p.c.}
\label{eq:n59}
\end{equation}
At second order
\begin{eqnarray}
W^+_{f,2}[m,A] &=&
-\frac{1}{8\pi}\tr\langle 0|
\Bigg[\Bigg(
\frac{1}{2}\left(\frac{2}{\beta}\right)^2
\phi_1(\frac{\beta}{2}(m-D_{01}))
-\frac{2}{\beta}\phi_0(\frac{\beta}{2}(m-D_{01}))
\left(\frac{1}{4}(D_{02}-D_{01})+\frac{m}{2}\right)
\nonumber \\ &&
-\int_m^{\infty}dt\,
	\tanh(\frac{\beta}{2}(t-D_{01}))
	\frac{\left(\frac{1}{4}(D_{02}-D_{01})^2+m^2\right)}{2t+D_{02}-D_{01}}
\Bigg) \frac{1}{(D_{02}-D_{01})^3}\bfE^2
\Bigg] |0\rangle 
\nonumber \\ &&
+ X_{12}+\hbox{p.p.c.}
\label{eq:new.84b}
\end{eqnarray}
In these formulas, p.p.c. means pseudo-parity conjugate, $D_0\to
-D_0$, and $X_{12}$ means the same expression exchanging the labels 1
and 2.

The functions $\phi_n(z)$ are given by
\begin{equation}
\phi_n(\omega)=  P_{n+1}(\omega)
-\int_\omega^{+\infty}
dz\frac{(\omega-z)^n}{n!}\left(\tanh(z)-1\right)\,,
\quad n=0,1,2,\dots \,,
\quad {\rm Re}(\omega)>0,
\label{eq:n61}
\end{equation}
where the integration path runs parallel to the real positive axis
towards $+\infty$. The $ P_n(\omega)$ are polynomials of degree $n$
\begin{eqnarray}
P_1(\omega) &=& \omega \,, \nonumber \\ 
P_2(\omega) &=&
\frac{1}{2}\omega^2-\frac{1}{6}\left(\frac{i\pi}{2}\right)^2 \,,
\nonumber \\
 P_3(\omega) &=&
\frac{1}{6}\omega^3-\frac{1}{6}\left(\frac{i\pi}{2}\right)^2\omega \,.
\end{eqnarray}
These formulas for $\phi_n(\omega)$ refer to ${\rm Re}(\omega)>0$. When
${\rm Re}(\omega)<0$ the property
$\phi_n(-\omega)=(-1)^n\phi_n(\omega)$ can be used.

\subsection{The effective action}
\label{sec:3.2}

We can now use \eq{12} to obtain the effective action of the
$2+1$-dimensional scalar field. Up to two Lorentz indices, $C[m,A]$
equals $C_2[m,A]$ which is given in \eq{n43}. $W^+_f[m,A]$ is given in
eqs. (\ref{eq:n59}) and (\ref{eq:new.84b}). In these latter
expressions the functions $\phi_n(z)$ are made explicit using
\eq{n61}, and the variables $\Omega$ and $\Dh_0$ are used instead of
$D_{01}$ and $D_{02}$ (cf. subsection \ref{subsec:II.C}). In addition,
we introduce a chemical potential by means of the replacement $A_0\to
A_0-\mu$ \cite{Landsman:1987uw} where $\mu$ is a real constant
c-number (recall that $A_0$ is antihermitian). This shift is gauge
invariant and it is equivalent to $\Omega\to e^{\beta\mu}\Omega$.

The effective action up to two spatial covariant derivatives at finite
temperature and density is thus
\begin{eqnarray}
W_{s,0}[m,A] &=& \int d^3x\,\tr \left[\varphi_0(e^{\beta\mu}\Omega) \right] \,,
\label{eq:n63}\\
W_{s,2}[m,A] &=& 
\int d^3x\,\tr \left[\varphi_2(e^{\beta\mu}\Omega_1,\Dh_{02})\bfE^2
\right] \,.
\label{eq:n64}
\end{eqnarray}

The functions $\varphi_0$ and $\varphi_2$ are given by
\begin{equation}
\varphi_0(\omega) = -\frac{1}{4\pi} \left(
\frac{1}{3}|m|^3+
\int_{|m|}^{+\infty} dt (t^2-m^2)
\frac{\omega}{e^{\beta t}-\omega}
 \right)
+{\rm p.p.c.} \label{eq:15b}
\end{equation}
\begin{eqnarray}
\varphi_2(\omega,y) &=&
-\frac{1}{8\pi} \frac{1}{y^3}
\Bigg[
\frac{1}{2}|m|y
+\frac{1}{2}\left(\frac{1}{4}y^2-m^2\right)
\log\left(\frac{2|m|+y}{2|m|-y}\right)
\nonumber \\ && 
+ \int_{|m|}^{\infty}dt\,\left(
	\frac{2\omega}{e^{\beta t}-\omega}
	\left(\frac{t^2-m^2}{2t-y}-\frac{1}{2}y\right)
	-\frac{2\omega}{e^{\beta (t+y)}-\omega}
	\left(\frac{t^2-m^2}{2t+y}+\frac{1}{2}y\right)
\right)
\Bigg]
\nonumber \\ && 
+ \hbox{p.p.c.}
\label{eq:13b}
\end{eqnarray}
In these formulas p.p.c. corresponds to $y\to-y$ and $\omega\to\omega^{-1}$.

The formulas (\ref{eq:n63}) and (\ref{eq:n64}), expanded in
eqs. (\ref{eq:15b}) and (\ref{eq:13b}), constitute the main result of
this section. They are necessarily complicated looking since an
infinite number of Feynman graphs (with any number of temporal gauge
fields, cf. Section \ref{subsec:IV.B}) are being added, and the
effective action is non-local in time at finite temperature. A much
simpler formula is presented below if only the lowest order is
retained in a further expansion in the number of temporal covariant
derivatives.

Some remarks are in order. In writing the formula we have already used
the fact that the effective action is an even function of $m$ (since
this is already true for $W^+_f[m,A]$). The condition ${\rm
Re}(\omega)>0$ in \eq{n61}, implies that the previous formulas refer
to the physically relevant case $|\mu|<|m|$ only.  All polynomial
contributions introduced by $P_n(\omega)$ in \eq{n61} combine in such
a way that all the polynomial dependence on $D_{01}$ cancels and the
result is a periodic function of $D_{01}$. This cancellation requires
the explicit and as well as the p.p.c. terms to hold and it is a
non-trivial check of the formulas. (Of course, the periodicity holds
also for the fermionic effective actions in the pseudo-parity even
sector, $W^+_f[m,A]$.) The argument of the logarithm in
$\varphi_2(\omega,y)$ is to be taken in the interval $(-\pi,\pi)$ and
$t$ runs on the real positive axis. (Note that $y$ represents $\Dh_0$
and thus it is purely imaginary.)

\subsection{Expansion in space-time derivatives at finite temperature}

A simpler expression for $W_{s,2}[m,A]$ is obtained retaining only the
leading order in an expansion in $\Dh_0$. This corresponds to expand
in powers of $y$ in the function $\varphi_2(\omega,y)$. As noted this
expansion does not break any symmetry and is a natural one in the
present context. This produces
\begin{equation}
W_{s,2}[m,A] = 
\int d^3x\,\tr \left[\varphi_{2,0}(e^{\beta\mu}\Omega)\bfE^2 \right]
 +O(\Dh_0)\,,
\label{eq:40b}
\end{equation}
with
\begin{eqnarray}
\varphi_{2,0}(\omega) &=&
-\frac{1}{96\pi}\frac{1}{m}
\frac{ e^{2\beta m}-\omega^2 -2\beta m e^{\beta m}\omega}
{(e^{\beta m}-\omega)^2}
+\hbox{p.p.c.}
\nonumber \\ 
&=&
-\frac{1}{96\pi}\frac{1}{m} \frac{d}{dm}\left[ m
\frac{ e^{\beta m}+\omega}{e^{\beta m}-\omega} 
\right]
+\hbox{p.p.c.}
\end{eqnarray}
Note that $\varphi_{2,0}(\omega)$ is an even function of $m$.  This
formula is manifestly invariant under all symmetries of the action.
Two non-trivial checks of the calculation are i) that negative powers
of $y$ have canceled and ii) that $\varphi_{2,0}(\omega)$ no longer contains
parametric integrals (i.e., $\int_{|m|}^\infty dt$). As noted, this
expression is exact for $W_{s,2}[m,A]$ in the Abelian and stationary
case.

\subsection{Zero temperature limit}
\label{subsec:zero-temp}

The zero temperature limit of eqs. (\ref{eq:15b}) and (\ref{eq:13b})
is straightforward: because $|\mu|<|m|$ and $\Omega$ is unitary, the
term $e^{\beta t}$ always dominates, thus the limit $\beta\to \infty$
just removes the terms with $\int_{|m|}^\infty dt$ and the dependence
in $\mu$ is also canceled (we are assuming $\mu$ fixed as $T\to
0$). That is
\begin{eqnarray}
W_{s,0}[m,A] &=& -\frac{1}{6\pi}\int d^3x\,\tr \left[|m|^3  \right]\,,
\qquad(T=0)
 \label{eq:15c} \\
W_{s,2}[m,A] &=&
-\frac{1}{8\pi}\int d^3x\,\tr
\Bigg[\bfE\Bigg(
|m|\Dh_0 
+\left(\frac{1}{4}\Dh_0^2-m^2\right)
\log\left(\frac{2|m|+\Dh_0}{2|m|-\Dh_0}\right)
\Bigg)\frac{1}{\Dh_0^3}\bfE\Bigg]\,.
\label{eq:13c}
\end{eqnarray}

At zero temperature Lorentz invariance is a symmetry of the
action. Such symmetry is not obvious in $W_{s,2}[m,A]$ since it sums
all orders in $\Dh_0$ but not in $\Dh_i$, $i=1,2$. However,
considering an expansion in inverse powers of $m$ (a Lorentz invariant
expansion) gives
\begin{eqnarray}
W_{s,2}[m,A] &=&
-\frac{1}{48\pi}\frac{1}{|m|}\int d^3x\,\tr
\left[\bfE^2\right] + O(m^{-3}) \,.
\label{eq:51}
\end{eqnarray}
The same result follows from \eq{40b}. This expression admits a
unique Lorentz invariant completion, namely
\begin{equation}
W_s[m,A] = -\int d^3x\,\tr \left[
\frac{1}{6\pi} |m|^3 
+\frac{1}{96\pi}\frac{1}{|m|}F_{\mu\nu}^2
\right] + O(m^{-3}) \,.
\label{eq:o5}
\end{equation}
This formula comes from a direct application of the results in
\cite{Chan:1986jq} thereby being a check for our formulas. At next
order in $1/m$ several Lorentz invariant operators of dimension 6 can
appear and \eq{13c} puts a constraint on their coefficients.  Note
that the last two formulas hold also at finite temperature since the
temperature-dependent corrections are $O(e^{-\beta|m|})$ (see also
Section \ref{subsec:IV.C}.)

\subsection{The partition function}

The effective action at finite temperature and density is directly
related to the grand-canonical potential, namely
$W_s[m,A;\beta,\mu]=\beta\Omega(\beta,\mu)$.  (Note that $\mu$
introduced by the replacement $\partial_0\to \partial_0-\mu$ couples
to the charge and not to the number of particles which is not
conserved in the relativistic case \cite{Haber:1981fg}). However,
strictly speaking, a system at equilibrium with temperature $T$ and
chemical potential $\mu$ can only be stationary. In addition, the
physical effect of a (negative) constant external scalar potential is
indistinguishable from that of a (positive) chemical potential, since
both add to the energy for positive charges and subtract for negative
ones, thus for the partition function $A_0$ should not be Wick
rotated, and $A_0$ is real instead of imaginary. All expressions
depend only on the combination $A_0-\mu$. (It would be algebraically
inconsistent not to rotate $A_0$ to its Euclidean version in the
general case, but not within the subset of stationary configurations.)
Note that the fact that $A_0$ is real or imaginary does not affect the
functional form; in any case the functional derivative of the
effective action with respect to $A_0$ yields the charge density and
the derivative with respect to $\mu$ yields the total charge:
\begin{equation}
\rho(x)= \frac{\delta W_s}{\delta A_0(x)}\,,
\qquad
Q= -\frac{1}{\beta}\frac{\partial W_s}{\partial \mu}\,.
\end{equation}

In view of the relation between partition function and effective
action, it follows that $W_{s,0}$ describes a two-dimensional
relativistic ideal gas in presence of an almost space-time constant
scalar potential $A_0$. An explicit calculation of the charge density
using $W_{s,0}$ in \eq{15b} and integrating by parts yields
\begin{equation}
\rho_0= \frac{N_f}{4\pi}\int_{|m|}^\infty dt\,
t\left(\coth(\frac{\beta}{2}(t-\mu))- \coth(\frac{\beta}{2}(t+\mu))\right) \,,
\end{equation}
$N_f=\tr(1)$ is the number of flavors. We are assuming that $\mu$
couples equally to all flavors and we have dropped $A_0$ since it can
be recovered from $\mu$.  This formula can rewritten in the standard
form \cite{Haber:1981fg}
\begin{equation}
\rho_0= \int \frac{d^2k}{(2\pi)^2}\left(
\frac{N_f}{e^{\beta(\omega(k)-\mu)}-1}
-\frac{N_f}{e^{\beta(\omega(k)+\mu)}-1}
\right) \,,
\end{equation}
where $\omega(k)= \sqrt{k^2+m^2}$.

Likewise, the term $W_{s,2}$ adds a contribution to the density
coupled to $A_0$. Let us consider the Abelian case (in addition to
stationary), then $\Dh_0$ vanishes and \eq{40b} becomes exact. There
will be two contributions to the charge density, one coming from the
explicit dependence on $A_0$ and another though the dependence in
$\bfE$. The latter is a total derivative and does not contribute to the
total charge. Both contribution can be combined to give
\begin{eqnarray}
\rho_2(x) &=& -\frac{N_f}{96\pi }\frac{1}{m}\frac{d}{dm}\Bigg( m\Bigg[
-\frac{\beta}{2}{\rm cosech}^2(\frac{\beta}{2}(m-\mu+A_0))\bfE^2
\nonumber \\
&& + 2\coth(\frac{\beta}{2}(m-\mu+A_0))\bfnabla\bfE \Bigg]\Bigg)
-\hbox{p.p.c.}
\end{eqnarray}
(Note that, $A_0$, $\mu$ and $\bfE$ are pseudoparity odd.) At zero
temperature (assuming $|m|>|\mu-A_0|$, or else at finite temperature
but large mass) this simplifies to
\begin{equation}
\rho_2(x)= -\frac{N_f }{24\pi }\frac{1}{|m|}\bfnabla\bfE \,,
\end{equation}
which also follows from \eq{51}.

It is also interesting to note the relation of our results with other
formulas using polylogarithms.\footnote{The polylogarithms are defined
as\cite{Lewin:1981}
$$\Li_n(z)= \sum_{k=1}^\infty\frac{z^k}{k^n}\,.$$} The functions
$\phi_n(\omega)$ introduced in \cite{Salcedo:1999sv} and subsection
\ref{sec:3} are directly related to polylogarithms, namely
\begin{equation}
\phi_n(\omega)= P_{n+1}(\omega)
-(-2)^{-n}\Li_{n+1}(-e^{-2\omega}) \,.
\end{equation}
This relation is easily established from \eq{n61} by noting that it
verifies the following defining properties of $\Li_n(z)$:
\begin{equation}
\Li_0(z)= \frac{z}{1-z}\,,\quad
z\frac{d}{dz}\Li_n(z)= \Li_{n-1}(z)\,,\quad
\Li_n(0)=0\,.
\end{equation}
In this notation, $W_{s,0}$ in eqs. (\ref{eq:n63}) and (\ref{eq:15b}) becomes
\begin{equation}
W_{s,0}[m,A] = -\frac{1}{4\pi}\int d^3x\,\tr \left[
\frac{|m|^3}{3}+2T^3\left(\beta|m|\Li_2(z) 
+\Li_3(z) \right)
 \right] +{\rm p.p.c.}
\end{equation}
where $z=e^{-\beta(|m|-\mu+A_0)}$ is the fugacity, and the
corresponding density becomes
\begin{equation}
\rho_0 = \frac{N_f T^2}{2\pi}
\left(\beta|m|\Li_1(z) +\Li_2(z) \right) -{\rm p.p.c.}
\end{equation}
in agreement with\cite{Blas:1999jn}.

As a final comment related to the partition function, we note that the
relation \eq{12} can also be written using the chemical potential
instead of the scalar potential $A_0$. That is, if
$\Omega_b(\beta,\mu)$ and $\Omega_f(\beta,\mu)$ represent the
grand-canonical potentials of a system of non interacting particles in
presence of external fields, treated as bosons or fermions
respectively, then
\begin{equation}
\Omega_f(\beta,\mu)= -\Omega_b(\beta,\mu+\frac{i\pi}{\beta})\,.
\label{eq:55}
\end{equation}
The minus sign comes because the functional integral with Grassman
variables gives the determinant of the quadratic form instead of the
inverse determinant. The shift $\mu\to \mu + \frac{i\pi}{\beta}$
accounts for the different boundary conditions. Of course a shift
$\mu\to \mu+\frac{2\pi i}{\beta}$ must leave the partition function
invariant, since $\mu$ is coupled to an integer-quantized charge. This
is another manifestation of gauge invariance. For interacting
particles \eq{55} can be extended using the well-known prescription of
adding a minus sign for each particle loop (\eq{55} corresponds to the
particular case of one-loop).\footnote{An equivalent procedure would
be to compute the grand-canonical potential for $n$ replicas of the
particles and then set $n$ to $-1$ at the end. That is, in the
notation of \cite{Negele:1988} the Hamiltonian becomes
\begin{equation}
H=\sum_{\alpha,\beta}h_{\alpha\beta}\sum_{\sigma=1}^na^\dagger_{\alpha\sigma} 
a_{\beta\sigma}
+ \sum_{\alpha,\beta,\gamma,\delta}v_{\alpha\beta\gamma\delta}
\sum_{\sigma_1,\sigma_2=1}^na^\dagger_{\alpha\sigma_1} 
a^\dagger_{\beta\sigma_2} 
a_{\delta\sigma_2} a_{\gamma\sigma_1} \,.
\end{equation}
}

\section{Further general remarks on the method}

\subsection{Remarks on gauge invariance}
\label{subsec:IV.A}

As noted in the introduction, topologically large gauge
transformations have played a prominent role in the development of
this field by putting severe constraints on the allowable forms of the
effective action functional. On the other hand it seems that in our
present approach such a role is played instead by the periodicity
constraint which prompts the appearance of the Polyakov loop
$\Omega(x)$. In this subsection we will make some remarks to try to
clarify the relation between both concepts.

Let $G$ be the gauge group, and let $\Gamma$ be the relevant homotopy
group controlling the existence of topologically large gauge
transformations at finite temperature.  In $0+1$-dimensions
$\Gamma=\pi_1(G)$, therefore there are large gauge transformations for
the Abelian group U(1) but not for simply connected groups such as
SU($n$). In $d+1$-dimensions (with $d$ different from zero), let us
assume for this discussion that the spatial boundary conditions are
such that the space becomes effectively compactified to a
sphere. Hence $\Gamma$ contains classes of mappings from the
space-time manifold ${\rm S}^1\times {\rm S}^d$ into the gauge group
$G$, where the factor ${\rm S}^1$ corresponds to the compactified
Euclidean time and ${\rm S}^d$ to the $d$-dimensional space. (Note
that, unless otherwise stated, we will require all functions to be
continuous on the space-time manifold, and in particular, periodic as
a function of time.) Theorem 4.4 of \cite{Whitehead:1995} then implies
that\footnote{When $d=0$ the theorem is consistent with
$\Gamma=\pi_1(G)\times\pi_1(G)$ which follows from ${\rm
S}^0=\{1,-1\}$. Physically, we want the spatial manifold at $d=0$ to
be just \{1\} and so $\Gamma=\pi_1(G)$.}
\begin{equation}
\Gamma/\pi_{d+1}(G)=\pi_1(G)\times\pi_d(G)\,.
\end{equation}
Let us consider the case $d=2$. Because the group $\pi_2(G)$ is always
trivial for any (compact) Lie group $G$, this simplifies to
$\Gamma/\pi_3(G)=\pi_1(G)$. Two typical cases are
\begin{itemize}
\item[i)] $G={\rm U}(1)$. In this case $\pi_3$ is trivial and $\pi_1$
is ${\mathbb{Z}}$, so $\Gamma={\mathbb{Z}}$. There are non-trivial
gauge transformations which wind $n$ times around the ${\rm S}^1$
factor of the space-time. They can be realized by space-independent,
but time-dependent, gauge transformations.
\item[ii)] $G={\rm SU}(n)$ ($n\ge 2$). In this case $\pi_1$ is trivial
but $\pi_3={\mathbb{Z}}$ and so once again $\Gamma={\mathbb{Z}}$. In this case
the corresponding large gauge transformations must be space-time
dependent.
\end{itemize}
Note that time-independent gauge transformations are controlled by the
homotopy group $\pi_2(G)$ which is trivial, and so they are always
topologically small in two spatial dimensions.

Let us now turn to the point of view used in this work. The
periodicity constraint refers to the fact that a gauge invariant
functional must depend on $\Omega(x)$ and not just on
$\log(\Omega(x))$. The cleanest way to formalize this is by working on
the gauge $\partial_0A_0=0$, in which $\Omega=\exp(-\beta
A_0(\bfx))$. Thus both $\Omega$ and $A_0$ are time-independent in this
gauge. Let us remark that taking $A_0$ to be time-independent is not
merely a restriction on the set of possible gauge field
configurations, it is a choice of gauge in the sense that every gauge
field configuration admits a gauge transformed configuration which is
$A_0$-stationary \cite{Salcedo:1999sv}. In this gauge the periodicity
constraint expresses that a gauge invariant functional must be a
periodic functional of $A_0$. To be concrete consider a generic
expression of the form (cf. \eq{n36a})
\begin{equation}
\Gamma[M,A]= \sum_n\int d^{d+1}x\,\tr[g_n(A_0){\cal O}_n]\,,
\label{eq:n60}
\end{equation}
where ${\cal O}_n$ are gauge covariant local operators, then gauge
invariance requires the functions $g_n(z)$ to be periodic with period
$2\pi i/\beta$. The necessity of this requirement follows immediately
from considering the following class of gauge transformations
\begin{equation}
U(x)= \exp(x_0\Lambda(\bfx))\,,
\label{eq:9}
\end{equation}
where the time-independent function $\Lambda(\bfx)$ takes values on
the Lie algebra of $G$ and is restricted by the following conditions
\begin{equation}
[A_0(\bfx),\Lambda(\bfx)]=0\,,\quad \exp(\beta \Lambda(\bfx))=1\,.
\label{eq:A3}
\end{equation}
The second condition means that the eigenvalues of $\Lambda(\bfx)$ are
of the form $\lambda_j=2\pi in_j/\beta$, for integer $n_j$ (such
integers are ${\bfx}$-independent by continuity) and it ensures that
the corresponding $U(x)$ is periodic in the temporal direction. Under
such a gauge transformation
\begin{equation}
A^U_0(\bfx)= A_0(\bfx) + \Lambda(\bfx) \,,
\end{equation}
i.e. the spectrum of $A_0$ is shifted by multiples of $2\pi i/\beta$
and the functions $g_n(z)$ must be periodic.

Before proceeding, an important point should be noted regarding the
approach used in this work. Namely, at the price of working with
asymptotic expansions, we can afford to derive formulas which are
``universal'' in the sense that no restriction is put on the algebraic
properties of the internal space, in particular, the formulas must
hold for any gauge group. This means for instance that the functions
$g_n(z)$ above are the same for all theories and configurations. The
requirement of universality puts stronger constraints on the
functionals that cannot be appreciated when working with concrete
theories only (for instance, in particular theories some of the
operators ${\cal O}_n$ can vanish identically and so the corresponding
function $g_n(z)$ does not play a role).

The transformations introduced in \eq{9}, subjected to the conditions
\eq{A3}, have been named discrete transformations associated to
$A_0(\bfx)$ in \cite{Salcedo:1999sv} since in general they form a
discrete set due to the condition $\lambda_j=2\pi in_j/\beta$. Note
that this set depends on the particular time-independent field $A_0$,
through the condition $[A_0,\Lambda]=0$. Clearly, these
transformations leave invariant the gauge condition
$\partial_0A_0=0$. Likewise, the gauge condition is also preserved by
time-independent gauge transformations. In \cite{Salcedo:1999sv} it is
proven that these two kinds of transformations are the most general
ones which preserve the $A_0$-stationary condition.\footnote{This is
the generic case which we will assume. It holds whenever $\exp(\beta
A_0(\bfx))$ is either nowhere degenerated or at least the regions of
degeneracy are sufficiently small that a unique eigenbasis can be
selected (up to normalization) by continuity \cite{Salcedo:1999sv}.
In this case $A_0(\bfx)$ is also nowhere degenerated and thus
$\Lambda(\bfx)$ is completely determined by its eigenvalues. So
generically the discrete transformations form a discrete set.}  This
means that within this gauge a functional such as $\Gamma[M,A]$ above
is gauge invariant if and only if the functions $g_n(z)$ are
periodic. Then it can be written in a manifestly gauge invariant form
(without gauge fixing) as in the right-hand side of \eq{n36a} with
$\varphi_n(e^{-\beta z})=g_n(z)$.  Note that invariance under
time-independent gauge transformations do not impose further
constraints on the $g_n(z)$.

The previous discussion suggests a comparison between large gauge
transformations, on the one hand, and discrete transformations, on the
other. Two questions pose themselves at this point. Are the discrete
transformations large in the topological sense? Is it necessary to
rely on large gauge invariance in order to arrive to an
$A_0$-stationary gauge?

The first question can be answered as follows
\cite{Salcedo:1999sv}. For a multiply-connected group such as U(1),
the non-trivial discrete transformations are topologically large since
they loop once or more on the temporal circle S$^1$. On the other
hand, for a simply-connected group such as SU($n$) (in more than one
space-time dimension) the discrete transformations associated to some
gauge configuration may be large or small, depending on $A_0(\bfx)$.
For instance if $A_0(\bfx)$ is everywhere diagonal, $\Lambda(\bfx)$ is
also diagonal and in fact a constant. In this case the discrete
transformation describes a single loop on the gauge group for all $x$
and it is homotopically trivial. In general, however, discrete
transformations can be topologically large.  It might seem that the
form of the discrete gauge transformations in \eq{9} factorizes time
and space and so it is always classified by the homotopy group
$\pi_1(G)$ being always small for a simply connected group. This is
true when $\Lambda(\bfx)$ is constant or homotopic to a constant, but
in general this is not the case. The reason is that although
$\Lambda(\bfx)$ is a map from S$^2$ into the Lie algebra of $G$ (a
vector space, and thus contractile), it cannot be contracted to a
point since the spectrum of $\Lambda(\bfx)$ is constrained to be in
$\frac{2\pi i}{\beta}{\mathbb{Z}}$.  An explicit SU(2) example in
$2+1$ dimensions is provided in\cite{Salcedo:1999sv}, namely,
\begin{equation}
U(x)= \exp\left(\frac{2\pi nx_0}{\beta}
i\bftau\bfx\right)\,,\quad n\in {\mathbb{Z}}\,,
\label{eq:63}
\end{equation}
where $\bftau$ are the Pauli matrices and $\bfx$ lies on the unit
sphere S$^2$ in ${\mathbb{R}}^3$. $U(x)$ covers SU(2) $2n$ times and
thus it is homotopically large for non-vanishing $n$.

Regarding the second question, whether a given gauge configuration can
be brought to a $A_0$-stationary gauge using only small
transformations, it also depends on the group and the configuration
\cite{Salcedo:1999sv}. For the group U(1), any gauge configuration is
in the same homotopy class as one which is $A_0$-stationary. For a
simply-connected group such as SU($n$), it depends on the initial
$A_0(x)$.\footnote{This can be seen as follows. Let $A_0(\bfx)$ be
some $A_0$-stationary configuration for which all its discrete
transformations are small, and let us further assume that
time-independent transformations are also small (for instance
$d=2$). It follows that all $A_0$-stationary configurations related to
the previous one by a gauge transformation are in the same homotopy
class. Thus, if $A^U_0(x)$ is a gauge transformed configuration with
$U(x)$ large, no small transformation will bring it to the
$A_0$-stationary gauge.}

An apparent paradox arises here. As emphasized in
\cite{Deser:1997nv,Deser:1998gp}, although perturbation theory breaks
large gauge invariance, it respects invariance under small
transformations, or equivalent, under infinitesimal ones. (Actually,
this is strictly correct for Abelian theories only. In non-Abelian
theories infinitesimal gauge transformations mix different orders,
however, the mixing is mild since it only involves finite sets of
Feynman graphs.) On the other hand, we have just seen that expanding a
functional such as $\Gamma[M,A]$ in powers of $A_0$ (which is a
perturbative expansion) destroys periodicity and thus gauge invariance
under discrete transformations. This looks paradoxical since the
breaking occurs even if the discrete transformations are topologically
small, and no breaking was expected in this case.

The resolution comes from the observation that the right-hand side of
\eq{n60} refers solely to configurations in the gauge
$\partial_0A_0=0$. Within this gauge the only allowed infinitesimal
transformations are the time-independent ones, for which no breaking
occurs. Non trivial discrete gauge transformations, even small ones,
cannot be reached continuously within the $A_0$-stationary gauge. In
this sense they are always topologically large. It should be realized
that the concept of homotopically trivial is a relative one. A
transformation which is topologically large within the gauge group $G$
can become small if $G$ is regarded as a subgroup of a larger group
$G'$ and deformations within $G'$ are allowed. Conversely, a small
discrete gauge transformation becomes homotopically non-trivial if one
insists on preserving the gauge condition $\partial_0A_0=0$. The
usefulness of the concept of small and large transformations remains:
perturbation theory provides a functional valid in the region of small
fields, this region is preserved by infinitesimal transformations and
thus the perturbative functional must be small gauge invariant. Large
transformations, on the other hand, necessarily move the gauge
configuration away from the perturbative region and thus the response
of the perturbative functional under large transformations is not
trustworthy. As a consequence invariance under large transformations
provides useful non-perturbative information and puts non-trivial
constraints on the functional. This holds whether the transformations
are large from the point of view of $G$ or from the point of view of
the submanifold of $A_0$-stationary configurations. 

For another argument, we can recall our previous remark that a
functional such as $\Gamma[M,A]$ in \eq{n60} must hold for all
theories at the same time. It is not surprising to find a breaking of
gauge invariance under discrete transformations in a perturbative
expansion, when such transformations happen to be large, and conclude
that non-trivial non-perturbative conditions, namely, periodicity,
are required to avoid the breaking. However, as we have emphasized,
the property of being topologically large or small depends on the
group, whereas the formula must hold in all cases, and so periodicity
must follow in all cases too.

In this subsection we have considered a gauge fixing condition in
order to deal with the quantity $\log\Omega$ in a simple way. We must
recall, however, that the expressions derived with the method studied
in this work are all fully gauge invariant, provided some regularity
conditions are met, since they depend on $\Omega$. This does not mean
that large or discrete gauge transformations play no role
whatsoever. This is because the regularity conditions fail for the
ultraviolet divergent pieces of the effective action. This translates
into the fact that the functions $\varphi_n(\Omega)$ (cf. \eq{n36a}), can
be many-valued (although $\Omega=0$ needs not be one of the branching
points). A typical example is the effective action of $0+1$
dimensional fermions, \eq{o1}. Further let us consider the Abelian
U(1) case, which admits large transformations and they coincide with
the non-trivial discrete transformations in the $A_0$-stationary
gauge. Choosing $\eta=+1$ the branching point is at
$\Omega=-\exp(\beta m)$. Under a large transformation, the argument
may go into a different Riemann sheet (depending on the sign of $m$),
yet the functional is such that it changes, at most, by an integer
multiple of $2\pi i$, reflecting the fact that the partition
functional is one-valued. This is a general property which puts
constraints on the functions $\varphi_n(\Omega)$.

Perhaps this is a good place to remark that the gauge invariance
implied by the use of the $\zeta$-function prescription, \eq{new.10a},
has two levels \cite{Salcedo:1999sv}. One the one hand, there is the
somewhat trivial gauge invariance implied by the fact that the
regularization depends solely on the spectrum. Since the spectrum of
the operator $K$ is left unchanged by gauge (or more generally
similarity) transformations, gauge invariance follows. However, the
definition of the $\zeta$-function introduces a branch cut in the
manifold of operators $K$, each singular operators being a branching
point on such a manifold. On a given Riemann sheet, the effective
action functional may display a jump discontinuity along the branch
cut. Because the determinant of $K$, being the regularized product of
eigenvalues, is a smooth functional, it follows that the jump must be
an integer multiple of $2\pi i$. This is a tighter constraint on top
of the trivial gauge invariance noted above. For instance, we have
noted in the introduction that perturbation theory for fermions at
finite temperature yields a Chern-Simons term which is renormalized by
a temperature dependent coefficient. Under large gauge transformations
this would introduce an unacceptable change in the effective action by
a quantity which is not a multiple of $2\pi i$. It would be tempting
to ``solve'' the problem by simply replacing the Chern-Simons term by
a suitable gauge invariant version of it, namely, the
$\eta$-invariant. This prescription restores gauge invariance but
introduces jumps which again are proportional to the temperature
dependent coefficient and thus it can be ruled out. The exact result
known in particular but non-trivial cases
\cite{Dunne:1997yb,Deser:1997nv,Fosco:1997ei} shows that this is not
the correct mechanism and that the determinant is a continuous
functional, with no jumps.

\subsection{Feynman graphs and large gauge invariance}
\label{subsec:IV.B}

We have already noted in subsection \ref{sec:3.2} that expressions
such as those in eqs.~(\ref{eq:n63},\ref{eq:n64}) involve an infinite
number of Feynman graphs. It seems interesting to understand which
Feynman graphs are being added and gain some insight on how
preservation of full gauge invariance is related to this.

To this end, we will first consider the simpler case of
$0+1$-dimensional fermions\cite{Dunne:1997yb}. The corresponding exact
effective action is given in \eq{o1} and that formula holds for
arbitrary gauge fields which need not be Abelian nor stationary. As
noted, when the gauge group is not simply connected it supports large
gauge transformations, which augment the value of the effective action
by $2\pi i kn$, $n,k\in{\mathbb{Z}}$, where $n$ is the winding number
of the gauge transformation and $k$ depends on the gauge group and the
sign of the mass (in higher spatial dimensions it may also depend on
the homotopic class of the gauge field configuration).

Because the $0+1$-dimensional formula is exact, all Feynman graphs are
included in this case. This suggests that all graphs are required in
order to reconstruct the Polyakov loop $\Omega$ appearing in the
formula. We conclude that large gauge invariance does not act
selecting a certain subset of graphs.  The same conclusion is
expected to hold in higher dimensional formulas since the dependence
on $\Omega$ there is qualitatively similar to that of the
one-dimensional case.

To further discuss this point let us restrict ourselves to the case of
an Abelian gauge group. In this case large gauge transformations
correspond to discrete shifts of $a=-\log\Omega=\int dx_0A_0$ and
\begin{equation}
W_f(a+2\pi in)= W_f(a)+2\pi i kn \,,
\label{eq:o2}
\end{equation}
where the integer constant $k$ is known. This equation contains all
the information on large gauge transformations, and it is completely
equivalent to the statement
\begin{equation}
W_f(a)= P(a)+ka, \qquad P(a+2\pi in)=P(a)\,.
\end{equation}
Therefore, large gauge invariance is equivalent to the strict
periodicity of the function $P(a)=W_f(a)-ka$. Feynman graphs
correspond to expand in powers of $a$, and looking for large gauge
invariance in terms of Feynman graphs corresponds to detect
periodicity of a function from its Taylor expansion. This seems to be
a difficult task.

A related issue is studying to what extent large gauge invariance
determines the effective action. In the previous $0+1$-dimensional
Abelian case we have seen that to comply with gauge invariance, $P(a)$
must be periodic, i.e.,
\begin{equation}
P(a)= \sum_{n\in{\mathbb{Z}}} c_n e^{na}\,,
\end{equation}
for some (infinite number of) coefficients $c_n$. No further
information can be extracted from gauge invariance, and the
coefficients $c_n$ are not determined.  In order to achieve further
restrictions on the function $P(a)$, more information has to be
provided. If, for instance, one knows that the partition function
$Z(a)=\exp(-W_f(a))$ (a periodic function) contains only a finite
number of periodic modes, the corresponding Fourier coefficients can
then be determined from a few perturbative terms.  This is actually
the case in $0+1$-dimensions \cite{Das:2000rc}, and can be traced back
to the fact that the corresponding Hamiltonian contains a finite
number of states, namely, the vacuum or the 1-fermion state
(cf. Section \ref{sec:3}). In higher dimensions, besides the number of
fermions, there is a momentum quantum number and a corresponding
kinetic energy contributing to the eigenvalues of the Hamiltonian,
thus in general the partition function will contain all kinds of
Fourier modes. To see how this works, it is sufficient to consider
fermions in the Abelian case with $A_i=0$ and $A_0$ a space-time
constant (such $A_0$ cannot be gauged away at finite temperature). Let
$\epsilon^0_k$ denote the single-particle levels of the Hamiltonian
when $A_0$ is set to zero (that is, the kinetic energy only) and let
$\epsilon_k$ be the levels when $A_0$ is switched on. Clearly,
$\epsilon_k=\epsilon^0_k+A_0$, and the label $k$ is related to the
momentum of the fermion. The partition function is
\begin{equation}
Z[m,a]=\prod_k(1+e^{-\beta\epsilon_k})=
\prod_k(1+\Omega e^{-\beta\epsilon^0_k})\,, \quad \Omega=e^{-a}\,.
\end{equation}
$Z[m,a]$ will be periodic in $a$ but with many periodic modes, unless
$d=0$. When $d=2$, integration over the label $k$ yields a particular
case of the formula for $W_{f,0}[m,A]$ in \eq{n59}.

In more than one dimension we have proposed an expansion in the number
of spatial covariant derivatives. The zeroth order (e.g. \eq{n63}),
which contains no spatial indices, corresponds to the sum of all
Feynman graphs with the spatial momenta $p_i$ (of the external gauge
fields) set to zero and no legs with spatial $A_i$, but any number of
$A_0$ external legs and the full dependence in the frequency
$p_0$. There are no odd-order terms in the expansion. The second order
(e.g. \eq{n64}) corresponds to all graphs with two $A_i$ or $p_i$,
that is, i) graphs with $p_i=0$ and two spatial gauge fields, plus ii)
graphs with one $A_i$ and $p_i$ kept up to first order in a Taylor
expansion of the Green functions, plus iii) graphs with no legs $A_i$
and $p_i$ kept up to second order in a Taylor expansion around zero
spatial momentum. Note that the method assumes that the space has
topology ${\mathbb{R}}^d$. The contribution of higher orders follows a
similar pattern. Eventually, all contributions, all Feynman graphs,
are added up.

The previous conclusions follow from inspection of the formulas or
else from making the expansion by introducing a bookkeeping parameter
in order to count spatial indices, as explained in Section
\ref{subsec:II.C}. For the simpler formulas obtained by further
expanding in powers of $\Dh_0$, (e.g. \eq{40b}), we have noted that
this expansion does not seem to follow from insertion of a bookkeeping
parameter and so different expansions can be obtained, all of them
being equivalent when added to all orders. The problem and its
interpretation in terms of Feynman graphs can be seen in a
particularly simple case: let us assume that the gauge group is
Abelian, that $A_0$ is a space-time constant and $A_i$ are
space-independent although time-dependent. In this case the Green
functions depend only on the frequencies $\kappa_n=2\pi in/\beta$ of
the fields $A_i$:
\begin{equation}
W= f_0(\Omega)+
\sum_n f_2(\Omega,\kappa_n)A_{i,n}A_{i,-n} +\cdots\,,\quad
A_i(x_0)=\frac{1}{\beta}\sum_ne^{\kappa_n x_0}A_{i,n}\,.
\end{equation}
In addition, $\Dh_0$ is equivalent to $\partial_0$ and so an expansion
in $\Dh_0$ is just a Taylor expansion in powers of $\kappa_n$, to be
made on top of the expansion in powers of $A_i$.  This gives the
interpretation in terms of Feynman graphs.  Because $\kappa_n$ is a
discrete variable, the function $f_2$ (and similarly for higher
orders) is only well-defined at those discrete values. The ambiguity
comes when it is smoothly extended to continuous values of $\kappa_n$
in order to carry out the Taylor expansion. Presumably this ambiguity
can be removed by using the choice suggested by Carlson's theorem.

\subsection{Large mass expansions}
\label{subsec:IV.C}

Large mass expansions of the effective action can be also considered,
as done in \cite{Deser:1998gp} for fermions using the heat-kernel
technique. Inspection of our formulas in \ref{sec:3.2} and
\ref{subsec:zero-temp}, show that as the mass becomes large the
temperature dependence is exponentially suppressed, being $O(e^{-\beta
|m|})$, thus the large mass expansion is an asymptotic expansion with
temperature independent coefficients. This is consistent with the fact
that the coefficients are local operators independent of the global
topology of the space-time manifold \cite{Deser:1998gp}. Therefore, in
order to carry out a large mass expansion one can simplify and start
with a zero temperature theory. The simplification of working at zero
temperature is enormous to the point that this problem can be
considered a solved one.  There is a very large body of work on this
subject, both for bosons and for fermions, largely summarized in
\cite{Ball:1989xg} and references therein.  Large mass expansions for
fermions with arbitrary Dirac operators and arbitrary dimension,
computed along the lines of the method discussed in this paper for
finite temperature, can be found in \cite{Salcedo:1996qy}. The large
mass expansion can also be used as a check of the finite temperature
formulas. In \ref{subsec:zero-temp} we have already noted that \eq{o5}
is consistent with the result more straightforwardly obtained from the
zero temperature method presented in \cite{Chan:1986jq}. Likewise, for
$2+1$-dimensional fermions, and starting from the full finite
temperature calculation, one finds \cite{Salcedo:1999sv}
\begin{eqnarray}
W^+ &=& -\frac{1}{48\pi}\frac{1}{|m|}\int d^3x\,\tr\,(
F_{\mu\nu}^2) + O\left(\frac{1}{m^3}\right)\,,
 \\
W^- &=& \eta\sigma\Theta(-\sigma m)W_{\rm CS}
-\frac{i\eta}{8\pi}\frac{\varepsilon(m)}{12m^2}\int d^3x
\epsilon_{\mu\nu\alpha}\tr(F_{\beta\mu}\Dh_\alpha F_{\beta\nu})
+O\left(\frac{1}{m^3}\right)  \,. \nonumber 
\end{eqnarray}
for the pseudo-parity even and odd components, respectively.  (In
these formulas $W_{\rm CS}$ is the Chern-Simons term, $\Theta$ and
$\varepsilon$ denote the step and sign functions,respectively,
$\eta=\pm 1$ depends on the irreducible representation of the Dirac
gamma matrices taken and $\sigma=\pm 1$ distinguish the two possible
$\zeta$-function definitions of the effective action, depending on the
branch cut in the function $z^s$.)  These results also derive more
directly from the zero temperature formulas in
\cite{Salcedo:1996qy}. The term with $F_{\mu\nu}^2$ in $W^+$ is that
with $H_4$ in eq.~(4.11) of \cite{Deser:1998gp}, already noted there,
whereas the term with $F_{\beta\mu}\Dh_\alpha F_{\beta\nu}$ is that
with $P_5$ in eq.~(4.12) of the same reference.

\section{Summary and conclusions}\label{sec:4}

Our findings can be summarized as follows.

1. The one-loop effective action at finite temperature and density,
   for bosonic or fermionic particles in presence of background fields
   (both gauge and non-gauge) with arbitrary internal symmetry group
   and arbitrary space-time dependence, can be written as a sum (an
   asymptotic series in general) of terms ordered by the number of
   spatial Lorentz indices. Each term is well-defined from the
   effective action functional itself and is separately gauge
   invariant under all gauge transformations.

2. These terms are amenable to explicit computation using a
   combination of $\zeta$-function and symbols method. We have shown
   that this kind of calculations can be carried out preserving full
   gauge invariance throughout, without assuming particular internal
   symmetry groups or special space-time configurations for the
   background fields. The same arguments show that previous
   calculations done fixing the gauge through the condition
   $\partial_0A_0=0$ can be repeated lifting this condition, and this
   is equivalent to rewrite the final original result in a manifestly
   gauge invariant way.

3. A further gauge invariant expansion can be taken in the number of
   temporal covariant derivatives in the adjoint representation.
   Within this expansion, all ultraviolet finite terms and more
   generally, all terms not related to essential anomalies, can be
   written as a sum of gauge invariant local operators (i.e.,
   constructed with $\Dh_\mu= [D_\mu,\ ]$ and $M$) times a function of
   the field $\Omega(x)$, which also transforms locally under gauge
   transformations. For those terms containing anomalies, topological
   pieces and multivaluation, the effective action still looks local
   in terms of a suitable gauge covariant version of $\bfA(x)$, in
   addition to $M$, $\Dh_\mu$ and $\Omega$.

4. The method is explicitly applied to the problem of relativistic
   scalar particles in $2+1$ dimensions. The corresponding effective
   action is computed up to terms with two spatial Lorentz
   indices. The result is checked against the known result at zero
   temperature and also the known partition function of a relativistic
   Bose gas. The corrections to the density are also
   computed. Finally, a simple rule is noted relating the bosonic and
   fermionic versions of the grand-canonical potentials of ideal or
   interacting systems.

\section*{Acknowledgments}
This work was supported in part by funds provided by the Spanish
DGICYT grant no. PB98-1367 and Junta de Andaluc\'{\i}a grant
no. FQM-225.

\end{document}